\documentclass[manuscript]{aastex}
\newcommand{\mod}{{\sc m19r8e1.25ni56}}
\shorttitle{Hydrodynamical model of SNe II~-P}
\shortauthors{Bersten et al.}

\begin{document}

\title{Hydrodynamical models of Type II-Plateau Supernovae}

\author{Melina C. Bersten\altaffilmark{1,2}, Omar
    Benvenuto\altaffilmark{3,4} and Mario Hamuy \altaffilmark{1} }

\affil{\altaffilmark{1} Universidad de Chile, Departamento de Astronom\'{\i}a, 
  Casilla 36-D, Santiago, Chile.}

\affil{\altaffilmark{2} Institute for the Physics and Mathematics of the
Universe, University of Tokyo, Kashiwanoha 5-1-5, Kashiwa, Chiba 277-8583,
Japan.}

\affil{\altaffilmark{3}Facultad de Ciencias Astron\'omicas y Geof\'{\i}sicas, Universidad Nacional de
La Plata, Paseo del Bosque s/n (B1900FWA) La Plata, Argentina.} 

\affil{\altaffilmark{4} Instituto de Astrof\'{\i}sica de La Plata, IALP, CCT-CONICET-UNLP, Argentina.}

\email{melina@das.uchile.cl}

\begin{abstract}
\noindent We present bolometric light curves of 
Type II-plateau supernovae (SNe~II-P) obtained using a newly
developed, one-dimensional Lagrangian  hydrodynamic code with flux-limited
radiation diffusion.
Using our code we calculate the bolometric light curve and 
photospheric velocities of SN~1999em obtaining a remarkably good
agreement with observations despite the
simplifications used in our calculation. The physical parameters used
in our calculation are $E=1.25$ foe, $M= 19 \, M_\odot$, $R= 800
\, R_\odot$ and $M_{\mathrm {Ni}}=0.056 \, M_\odot$. We find that 
an extensive mixing of $^{56}$Ni is needed in order to reproduce a
plateau as flat as that shown by the observations. We also study the
possibility to fit the observations with lower values of the initial mass
consistently with upper limits that have been inferred from
pre-supernova imaging of SN~1999em in connection with stellar
evolution models.
 We cannot find a set of physical parameters that reproduce well the
observations for models with pre-supernova mass of $\leq$ 12
$M_\odot$, although models with 14 $M_\odot$ cannot be fully discarded.
\end{abstract}



\section{INTRODUCTION}
\label{sec:intro}

Type~II-Plateau SNe (SNe~II-P) form a well-defined family
characterized by a ``plateau'' in the optical light curve 
\citep{1979A&A....72..287B}, where
the luminosity remains nearly constant for a period of $\sim$100 days
, and the presence of prominent P-Cygni hydrogen
lines in the spectrum. They constitute a subclass of
the core-collapse SNe (CCSNe) ---which includes Type Ib,  Type Ic, and other
subclasses of type II SNe--- originated by the violent death of
stars with initial masses greater than 8 $M_\odot$
\citep{2003ApJ...591..288H,2009MNRAS.395.1409S} and sharing, in
general terms, the same explosion mechanism. It has been shown that
SNe~II-P are the most common type of SNe in nature, amounting to $\sim$ 60\%
of all CCSNe. Apart from their astrophysical importance in connection
with stellar evolution and the physics of the interstellar medium,
additional interest on SNe~II-P has recently arisen from the fact that
they have been established as good distance indicators with potential
application to cosmology, independent of Type Ia
SNe \citep{HP02}.

Massive stars may
suffer considerable mass loss during early phases of their evolution,
due to strong stellar winds or transfer to binary companions.
Thus, they may lose part or all of their outermost envelope of unprocessed
hydrogen and helium. Therefore, the vast diversity 
observed among CCSNe is related with the properties of the  
progenitor star. In the current picture, SNe~II possibly have the
least massive progenitors of all CCSNe subtypes which retain a significant
fraction of their external hydrogen layers.  This picture is
combined with hydrodynamical 
models of SNe~II-P,  which show that a red 
supergiant progenitor with an extensive H envelope is necessary in
order to reproduce the plateau-shaped light curves 
\citep{1971Ap&SS..10...28G,1977ApJS...33..515F,1976ApJ...207..872C}. Recent
direct detection of the progenitors of several  
SNe~II-P have confirmed this prediction
\citep{2003PASP..115.1289V,2004Sci...303..499S}.
There is general agreement that the explosion of  a massive star is originated
by the collapse of its central parts into a neutron star or black hole
when the iron core is formed at the end of the star evolution, and
further nuclear burning no longer provides thermal pressure to support
the star. However, the mechanism of the energy deposition in the
envelope still remains unknown in spite of the intensive theoretical
modeling  done in recent years \citep[see][and references
  therein]{2006NewAR..50..487B,2007PhR...442...38J}. The
approach usually followed to model supernovae is to decouple the
explosion in two independent parts: a) the core collapse and formation of
the shock wave, and b) the ejection of the envelope. Based on the analysis of
the propagation of the shock wave through the envelope, independently
of how the shock is formed, it is possible to study the observational
outcome of the explosion such as light curves and spectra. This
approach has been extensively used \citep[][among
  others]{1977ApJS...33..515F,
  1971Ap&SS..10...28G,1988ApJ...330..218W} and it has led to the conclusion that
the main factors influencing the outcome are the explosion
energy and the progenitor structure.

Observationally, SNe~II-P show a wide range 
of plateau luminosities ($L_p$) and durations ($\Delta t_p$),
expansion velocities ($v_{exp}$), and
nickel masses ($M_{\mathrm Ni}$) \citep{1989ApJ...342L..79Y,H01}. The
morphology of the light curve (LC) of SNe~II-P has been shown to be
connected with physical properties of the progenitor object such as 
ejected mass ($M$), explosion energy ($E$), and 
pre-supernova radius ($R$). The relations
between physical parameters and observables ($L_p$, $\Delta t_p$,
$v_{exp}$) were first derived analytically by
\citet{1980ApJ...237..541A} and 
then generalized by \citet{1993ApJ...414..712P}. Numerical
calibrations of these relations
were then  given by \citet{1983Ap&SS..89...89L,1985SvAL...11..145L} (LN83 and LN85
hereafter, respectively) based on a grid of hydrodynamical models for
different values of $M$, $R$ and $E$.  
\citet{2003ApJ...582..905H} and \citet{2003MNRAS.346...97N} applied 
such calibrations to a set of $\sim$ 20 SNe~II-P
observations and thereby derived masses,  
radii and explosion energies for their sample. However, these
studies have not been fully satisfactory. As was noted in
\citet{H01} and \citet{2009ApJ...701..200B}, some of the problems of inferring
physical quantities from the relations of LN83 and LN85 are: 1) the lack of
good-quality data, 2) the use of simplified relations between
ill-defined and hard-to-measure photometric and spectroscopic parameters, and 
3) the fact that some of the models are based on simplified physical
assumptions. Some of the  weaknesses of the LN83 and LN85 models are that
they did not include the effect of nickel heating in their calculations, the
use of old opacity tables, the neglect of any effect from 
line opacities, and the initial pre-supernova models adopted.
With the aim of  addressing these problems and gaining a better
knowledge of the  pre-supernova
properties of  SNe~II-P, we have 1) enlarged the data set of spectra
and light curves for 33 SNe~II-P \citep{H09}, and 2) developed our own
hydrodynamical model. This allows us to perform a comparison between
models and data in a consistent way.

The development of our hydrodynamics code started in 2006
and is part of MB's PhD thesis. In the interim since the
beginning of the work there have been important advances in the
field. For example, \citet{2007A&A...461..233U} performed a
detailed analysis of SN~1999em and also studied
how several physical parameters affect the LC. In that work, the
author provided relations between physical and observed parameters which
are valid only for SNe~II-P with similar properties to those of
SN~1999em.  In addition \citet{2009ApJ...703.2205K} calculated a 
set of modeled LC and spectra of SNe~II-P for different masses,
metallicities, and explosion energies, using initial models obtained from
stellar evolution calculations. They employed their models to describe
the dependence of plateau luminosity and duration on explosion
energy and progenitor mass. Nevertheless, the relations they found are
simple and easy to apply {\em only} in the extreme case of no
$^{56}$Ni production. When $^{56}$Ni is considered, the relations
involve more parameters, which hinders their applicability to obtain
physical parameters from observations. More recently,
  \citet{2010MNRAS.408..827D} studied the dependence  of the
properties of SN~II-P ejecta on explosion energy and pre-supernova stellar
evolution model. They found that the main-sequence mass
of SNe~II-P can be contrained using two measurements: (a) the
photospheric velocity at 15 days after shock breakout, which is a
good indicator of the explosion energy, and (b) the width of the nebular
phase O~{\sc i} $\lambda \lambda$6303-6363 \AA\ line  as an
indicator of the  helium-core mass and thereby of the mass of the star
in the main sequence. Other works have also  
explored the effect of several physical parameters on the LC and
other properties of SNe~II-P
\citep{2003MNRAS.345..111C,2004ApJ...617.1233Y}, although without
attempting to derive relations between  physical and observed
parameters. In  recent years, there has been a number of works
which analyzed the physical properties of individual SNe~II-P based
on hydrodynamical models \citep[][among others]
{2005AstL...31..429B,2009A&A...506..829U}.

In addition to the comparison with hydrodynamical models,
  there is an alternative way of deriving progenitor masses of
  SNe~II-P. That is, using pre-supernova
images to search for the progenitor star, and measuring its brightness
and color to derive a mass in connection with a stellar evolution
model. Currently, there are three SNe~II-P with a firm detection of
the progenitor and $\sim$ 17 without a positive 
detection  \citep{2009MNRAS.395.1409S}. The latter cases are still
useful since they provide  upper limits to the progenitor masses. 

At this stage there are three SNe~II-P with hydrodynamical masses which have
been studied in pre-supernova imaging (namely, SN~1999em, 
SN~2004et and SN~2005cs). As noted by
\citet{2008A&A...491..507U,2009arXiv0908.0700S},
in all cases the mass estimated by hydrodynamical models
is higher than the estimate or upper limit given by pre-supernova
imaging. This discrepancy poses a very 
interesting and unsolved problem which is necessary to study. 

In this work, we present our one-dimensional, flux-limited Lagrangian
hydrodynamical code which is used for modeling bolometric light curves of
SNe~II-P. In general the hydrodynamic modeling of SNe~II-P is easier
than the one of other CCSNe because the former explode in a very 
low-density environment
\citep{2000ApJ...545..444B,2006ApJ...641.1029C} and they 
possess extended,  nearly spherically symmetric hydrogen
envelopes which smooth out possible inhomogeneities arising from
differences in the explosion itself
\citep{1989ApJ...341..867C,2001PASP..113..920L}.  
In addition, at least during the optically thick plateau phase, we expect a
photosphere radiating as a ``dilute'' blackbody whose properties are mainly 
driven by the photospheric temperature \citep{E96,D05}. We employ the usual
hypothesis that the explosion can be decoupled into  
the collapse of the core and the ejection of the envelope. The energy
which is transferred to the envelope 
(denoted by us as ``explosion energy'') plays the role of a
coupling parameter between the internal and external problems. The
processes which control the envelope
ejection and the supernova radiation do not depend on how the energy
is transferred to the envelope as long as this process occurs during a
short enough time and in a small enough region near the
core. This means that for the
purpose of modeling light curves it is possible to assume the core as a
point-like mass and energy source
\citep{WW90,1999AstL...25..649D,2003MNRAS.345..111C}.   
Thus, our model is unable to describe details of the complex
explosion mechanism. 

The aim of developing our own code is to have 
available a simple, easy and fast model to study, in a consistent way,
the observed and physical parameters that determine the plateau phase
for a sample of 33 SNe II-P which will be presented in
a forthcoming paper. In this work we present a detailed
model for the prototypical SN~II-P 1999em, as test case to
show the consistency of the model. 
It will not be our goal in the future to craft specific models for each object
 but to present a general study of the properties of the whole sample
 of SNe~II-P.  

A description of our
numerical method, micro-physics and pre-supernova models 
is given in Section~\ref{sec:constitutive}. As a first step in our
comparison with data, in Section~\ref{sec:sn99em} we analyze the case
of the prototype SN~1999em, and we discuss how our model
compares with previous hydrodynamical studies of this object. In
Section~\ref{sec:lowmass} we test the feasibility of fitting the
observations of SN~1999em with low mass models, consistent with  
the upper limits obtained through pre-supernovae imaging in connection
with stellar evolution models. This
analysis also shows the sensitivity of our bolometric LCs on
the variation of physical parameters. Finally, in
Section~\ref{sec:conclusion} we give our conclusions.

\section{CALCULATION METHOD}
\label{sec:method}
Our supernova models are computed by numerical integration of the
hydrodynamical equations assuming spherical symmetry for a self
gravitating gas. The radiation 
transport is treated in the diffusion approximation 
with the flux-limited prescription of  \citet{1981ApJ...248..321L}. The
explosion is simulated by injecting a certain amount of  
energy during a very short time as compared with the hydrodynamic
time-scale  near the center of the progenitor object. This energy
induces the formation of a powerful shock wave that propagates through the 
progenitor transforming thermal and kinetic energy of the
matter into energy that can be radiated from the stellar surface. To
calculate shock waves, we include, as usual, an artificial-viscosity term in
the equations of moment and energy.

The equations and numerical method used are discussed in 
(\S~\ref{sec:equations}). In \S~\ref{sec:constitutive} we give a brief  
description of the constitutive relations used in our code. The energy
deposited by radioactive decay is discussed in \S~\ref{sec:gamma}. The
initial models are described in \S~\ref{sec:initialmodel}. A discussion
 of the  approximations used in our calculations is given in
 \S~\ref{sec:assumption}. Finally, in \S~\ref{sec:model_obs} 
we describe which parameters of the model are suitable to compare with
observations.

\subsection{Equations for radiation transport and hydrodynamics}
\label{sec:equations}
Our code follows a one-dimensional, Lagrangian prescription that solves
for the variables: radius ($r$), velocity ($u$),
density ($\rho$) and temperature ($T$) as a function of the Lagrangian
mass coordinate $m$, using the following equations which simply
express conservation laws:

\begin{mathletters}
\begin{equation}
\frac{\partial r}{\partial t}= u, 
\label{eq1} 
\end{equation}

\begin{equation}
V= \frac{1}{\rho}= \frac{4 \pi }{3} \frac{\partial r^3}{\partial
  m}, 
\label{eq2} 
\end{equation}

\begin{equation}
\frac{\partial u}{\partial t}= - 4 \pi r^2  \,\frac{\partial}
         {\partial m}(P+q) - \frac{G m} { r^2},  
\label{eq3}
\end{equation}

\begin{equation}
\frac{\partial E}{\partial t}= \epsilon_{\rm Ni} - 
      \frac{\partial L}{\partial m} - (P+q)  \, 
       \frac{\partial V} {\partial t},
  \label{eq4}
\end{equation}

\begin{equation}
     L = - ( 4 \pi r^2 ) ^2  \, \frac{\lambda ac}
     {3 \kappa} \, \frac{ \partial T^4} {\partial m}. 
  \label{eq5}
\end{equation}
\end{mathletters}
 
\noindent Here, $V$ is the specific volume, $P$ is the total pressure
(of gas and radiation), $q$ is the artificial viscosity which is included in the
equations to spread the pressure and energy over several mass zones
at the shock front. There are many expressions for the artificial viscosity,
all dependent on  the velocity gradient, which aim at providing a
convenient interpolation scheme between
unshocked and shocked fluid. We adopt the expression given by
\citet{VN50}. $E$ is the internal energy per unit of
mass (of gas and radiation), $\epsilon_{\rm Ni}$  is the energy
deposited  by the radioactive decay of nickel as we describe in \S
~\ref{sec:gamma}. We do not consider other sources of cooling or 
heating in equation~(\ref{eq4}), such as losses due to neutrino
processes or energy released by thermonuclear reactions. 
Even if neutrinos are very important in the 
formation of the shock wave as the explosion depends noticeably on the
efficiency of their energy transfer, most of them are 
emitted before the shock wave reaches the stellar photosphere, so they 
have no effect on later epochs of the SN evolution
\citep[see][]{Hill90,1991supe.conf..393B,2007PhR...442...38J}. The 
energy released by explosive nucleosynthesis is much less than the
energy of the shock wave as has been previously 
established [\citet[see][]{1965SvA.....8..664I,WW90,1996snih.book.....A}].
$T$ is the temperature of both matter and radiation, $\kappa$ is the
Rosseland-mean opacity and $L$ is the luminosity. Finally, $\lambda$
is the so-called 
``flux-limiter'', included in the equation of radiative transfer in
the diffusion approximation 
 to ensure a smooth transition between diffusion and
free-streaming regimes to assure causality. The expression 
adopted for $\lambda$ is,  

\begin{equation}
\lambda = \frac{ 6 + 3 R}{6 + 3 R + R^2}, 
\end{equation}

\noindent where  

\begin{equation}
R= \frac{ \mid \nabla T^4 \mid } {\kappa \rho \ T^4}= 
    \frac{4 \pi r^2}{\kappa T^4} \mid  \frac{ \partial T^4} {\partial m} \mid. 
\end{equation}

\noindent The quantities $E$, $P$, $q$ and $\kappa$ are functions of $\rho$,
$T$, and  chemical composition. Further details on these relations are
given in \S ~\ref{sec:constitutive}.

As boundary conditions we use $u=0$ near the center (at  $m=M_{\rm core}$,
where typically we adopt $M_{\rm core}= 1.4 \, M_{\odot}$), and 
$P_{\rm gas}=\rho= 0$ at the surface ($m=M$).

We discretize the previous equations using a space-centering
discretization with the extensive quantities evaluated  
at the interfaces and the intensive quantities in the midpoints of the
grid zones. Two time steps are adopted in each cycle:
one to advance the velocity, and the other 
to advance the material state variables. The time
step is chosen depending on limitations of stability and accuracy (see
the discussion below). Special care is taken in the centering of the
opacity used in the discretization of equation~(\ref{eq5}) in order to
prevent numerical noise to appear due to the propagation of the
radiation flow at the steep front where the opacity changes
significantly \citep{Christy67}.

We use an explicit scheme for the integration of the hydrodynamic
equations, but a semi-implicit scheme for the temperature, similar to the
one used by \citet{1977ApJS...33..515F}. The equations are linearized in 
$\delta T$ and solved iteratively for each time step using the
tridiagonal method. The discretization typically uses 300 mesh
points, with a finer sampling for the outer layers typically smaller
than $10^{-6} M_\odot$. This value was chosen based on tests performed
with our model and following previous works
\citep{1988ApJ...330..218W,1992ApJ...393..742E} which 
show that the early light curve is sensitive to the mass zoning in the
outer layers when a coarse grid  is used. 

Given that we are using an explicit hydrodynamic scheme, 
the time step should be chosen as a fraction of the minimum of the
Courant condition for all  zones in order to achieve
stability. We note that we have also imposed additional
  conditions on the time step: we have required that changes in temperature,
  density and flux over one time step be less than 5\%. 

The formation of the shock wave (SW) following core collapse is simulated by
artificially adding internal energy (``thermal bomb'') almost
instantaneously in the central region of the core. We usually
set a mass cut of $1.4 \, M_\odot$  which is the material assumed
to collapse and form a neutron star or black hole.
 Specifically, we employ an exponential
function both in mass and in time to distribute the injected energy across
several layers and some time steps, which helps to improve the numerical
treatment. With the scale factors used in this work for the
exponential functions almost all of the 
explosion energy was completely injected in a shell of 0.1 M$_\odot$
and within less than 1 minute, i.e. a short time compared with
the hydrodynamical time scale. Although there may be some differences
 in the velocity and temperature profiles during the shock
 propagation on the scale factors used. Our values were chosen such
 that the differences in the velocity profiles using this value or
 lower values were not significant.
 We have also tested explosions generated by 
injecting kinetic energy and we have obtained similar
results. The latter method, however, leads to very short 
numerical time steps, and therefore to slower calculations.
We have checked the accuracy of our calculation method by testing the
 conservation of  energy. The total amount of energy is
   conserved within $0.6$\% but if we consider the conservation of
   energy between two consecutive time steps, this 
   is even better,  within 4 $\times 10^{-6}$.  We consider it very
 acceptable for our purposes.

\subsection{Input physics}
\label{sec:constitutive}

The equation of  state (EOS) is  calculated using
simple expressions for $T$, $\rho$, composition,  and  for the
ionization degrees of hydrogen and helium corresponding to local
thermodynamical equilibrium (LTE). Therefore, the degree of ionization is 
determined by solving the corresponding set of Saha equations for
ionization of hydrogen and the first and second
ionization of helium. Ionization of heavy elements is
neglected in our EOS  (but, of course, it is taken
into account in the calculation of the opacity). The
 degeneracy pressure  is also included in the EOS. We tested our
results using a more sophisticated EOS, 
such as the one available from the Los Alamos
  Tables \citep{1996ApJ...456..902R}, and we did not find any significant 
differences with respect to the results obtained using our simple EOS.
This is expected because of the low densities attained in most of the
layers of the models. 

The Rosseland mean opacities, $\kappa$, used in our calculations are
derived using the OPAL opacity tables \citep[][and references
  therein]{1996ApJ...464..943I}. As these tables are given for $T > 6
\times 10^3$ K, we complement them with the opacity table provided by 
\citet{1994ApJ...437..879A} for lower temperatures which
includes molecular opacities. The tables are interpolated to each
other in order to guarantee a smooth transition at $T=10^4$ K.

These tables allow us to calculate opacities for several
metallicities. Also, for a fixed metallicity, 
different mixtures of H, He, C and O can be used. Although in this 
work we adopt a fixed value of $Z=0.02$, departures from this
value (as expected in the inner regions of the object) are
taken into account as excesses of C and O with respect to the values
of the adopted metallicity, at the expense of He.

\begin{figure}
\begin{center}
\includegraphics[scale=.40]{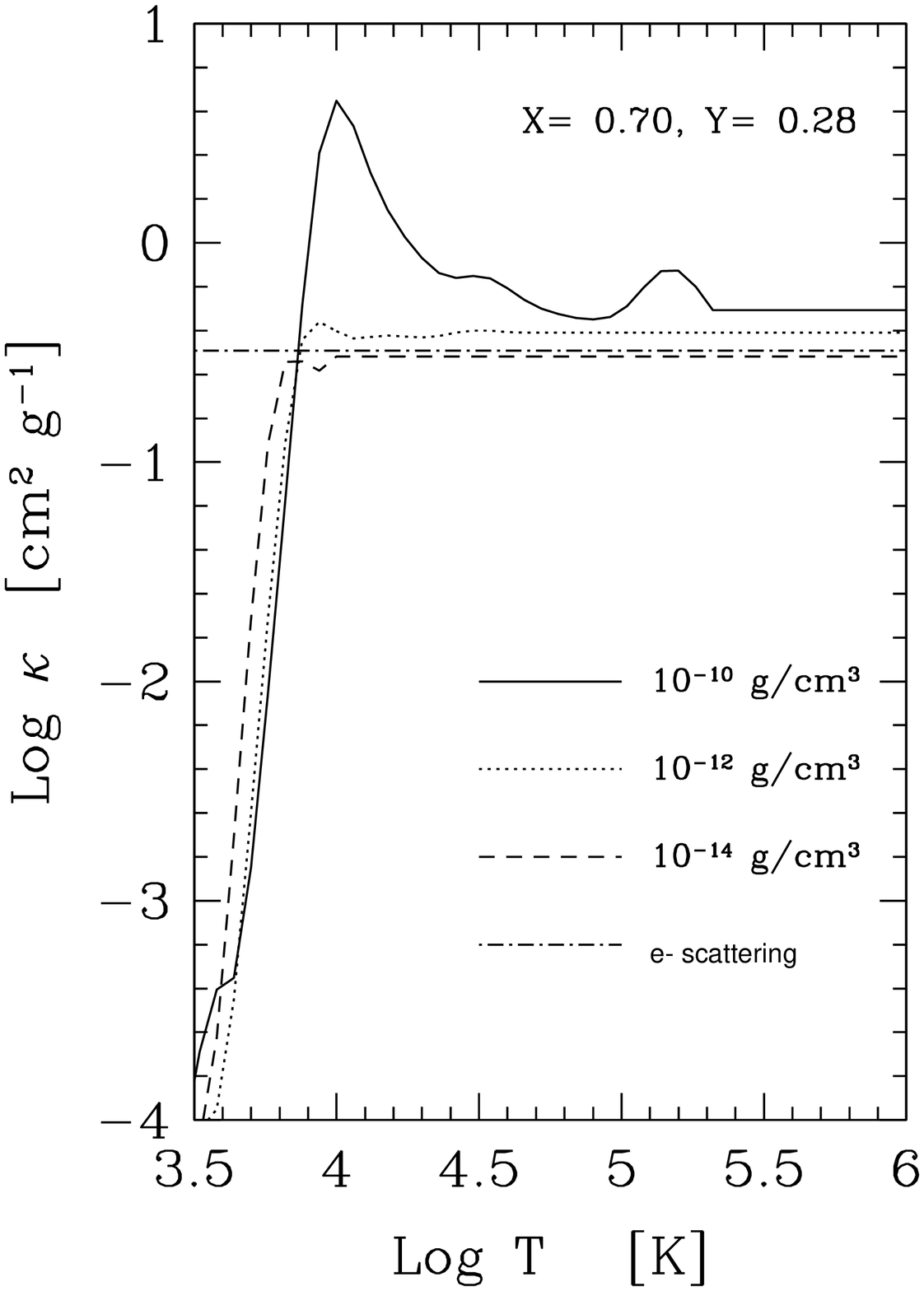}
\caption{The run of the Rosseland opacity on
  temperature ($T$) and density ($\rho$) for solar metallicity
  ($Z=0.02$) as used in our calculation without including any ``opacity
  floor'' (see discussion in Section~\ref{sec:constitutive}). 
  The ranges of $T$ and $\rho$
  shown are typical for SNe~II-P. The hydrogen (X) and  helium (Y)
  mass fractions used are 
  indicated. We also include the electron scattering
   opacity (considering full ionization, which is certainly
   unrealistic for the low temperature sector of 
  this plot) in order to show the dominance of this source for
   $T > 10^4$ K and  $\rho < 10^{-10}$ g cm$^{-3}$. Note that this
  value of the density is 
  reached early-on in the supernovae evolution. For lower 
  temperatures the absorption processes become an important source of
  opacity.\label{fig:opal}}  
\end{center}
\end{figure}

Figure~\ref{fig:opal} shows the opacity given by the tables as a
function of temperature for $Z=0.02$ and different densities. 
The range of temperature shown corresponds to the values reached
during the evolution of a SN~II-P while the values of density are
restricted to those achieved during the plateau phase.
Also shown in the plot is the contribution to the opacity from
electron scattering. Note that electron scattering is the dominant
source of opacity for $T >10^4$ K and $\rho < 10^{-10}$ gr cm$^{-3}$.

The Rosseland mean opacity includes scattering and absorption
processes. Scattering is the dominant process in the supernova ejecta
during the plateau phase when the densities are $\rho < 10^{-10}$ (see
Figure~\ref{fig:opal}), and this will be so until most of the
electrons recombine with ions and absorption processes become an
important source of opacity. On the other hand, 
in rapidly expanding envelopes where large velocity gradients are present,
the Rosseland mean opacity underestimates the true line opacity
\citep{1977ApJ...214..161K}, which hinders the estimation of the 
actual opacity in the outermost (recombined) layers.
Another effect that is not included in the calculation of $\kappa$ is
the non-thermal excitation or ionization of electrons which are
created by Compton 
scattering of $\gamma$-rays emitted by radioactive decay of $^{56}$Ni and
$^{56}$Co. The LTE ionization used in the calculation of
  $\kappa$ considerably underestimates the true ionization.
The correct way to treat these effects is to calculate the actual
contribution of non-thermal ionization to the opacity and to
  include the expansion opacity of 
lines. However, such treatment is beyond the scope of this paper. We
adopt an alternative approach to the problem that has been extensively used in
the literature
which consists in using a minimum value of the opacity (or ``opacity
floor'')  New: to partially solve the shortcoming in the Rosseland mean
  opacity. Given the dependence of the opacity on composition, usually
  two values of the opacity floor are used: one for the 
  H-rich envelope material and another for the metal-rich core. The opacity
  at each time and mesh point is chosen as the maximum value between 
  the tabulated Rosseland mean opacity and the opacity floor for the
  corresponding composition.

  Values of the opacity floor should be based on
  contributions to the opacity which are not included in the 
Rosseland mean opacity tables.  There are 
  differences in the values adopted in the literature.  For example,
  \citet{1990A&A...233..462H} adopted a value of 0.01 cm$^2$g$^{-1}$
  for the whole structure.
  \citet{1990ApJ...360..242S}  used  minimum values for
  the bound-free and bound-bound opacities of 
  k$_{bf}$= 9 $\times 10^{-3}$ Y cm$^2$g$^{-1}$ and k$_{bf}$= 1 $\times 10^{-2}$
  Z cm$^2$g$^{-1}$,  where Y and Z denote the mass concentration of
  helium and heavy 
  elements. Note that it is needed to add scattering and
  free-free absorption effects to the latter in order to obtain the Rosseland mean
  opacity. \citet{1991ApJ...374..266S} employed
  0.05 cm$^2$g$^{-1}$ for the envelope material, and 0.1 cm$^2$g$^{-1}$
  for the metal-rich core material. And \citet{2004ApJ...617.1233Y}
  used a value of  0.25 cm$^2$g$^{-1}$  for the helium-rich core, and
 0.01 cm$^2$g$^{-1}$ for the hydrogen-rich envelope. We do not intend
 to be exhaustive but the examples above provide a summary of the 
 values adopted in the field of SNe II.

 In this work, the minimum opacity values adopted are: $0.01$ cm$^2$ 
g$^{-1}$ for the envelope material, and $0.24$ cm$^2$g$^{-1}$ for the
metal-rich core material. These values were chosen based on a
comparision performed between our code and the STELLA code
\citep{1993A&A...273..106B,1998ApJ...496..454B} using the same initial
model provided by \citet{2005ApJ...619..427U}. Note that STELLA is an implicit
hydrodynamic code that incorporates multi-group radiative
transfer, and which additionally uses different opacity tables
  and includes the effect of lines opacities
  \citep{{2002nuas.conf...57S}. In spite of our approximations, it is
important to remark that we 
obtain an excellent overall agreement of the LC given by both
codes  (both in terms of the
duration of the plateau and of the morphology of the bolometric
LC) when the opacity floor is set the values given above.}

\subsection{Gamma-ray Deposition}
\label{sec:gamma}
The  radioactive decay of $^{56}$Ni $\rightarrow$
$^{56}$Co $\rightarrow$  $^{56}$Fe constitutes a relevant source of energy
which heats the envelope of the supernova. Several studies of
SNe~II-P have usually ignored the contribution of such component
during the early SN evolution and have only considered it beyond the end
of the plateau phase. Such approach would be correct  only if $^{56}$Ni were
very deeply  concentrated in the ejecta, in which 
case the local deposition of gamma rays is a good approximation because
 gamma photons can hardly diffuse out the regions where they are formed. 
Nevertheless, as was shown in studies of SN~1987A, there is no reason
to assume this type of $^{56}$Ni 
distribution,  \citep[][among others]
{1988A&A...196..141S,1988ApJ...324..466W,1988ApJ...331..377A,2000ApJ...532.1132B} 
and therefore we need to calculate the diffusion of gamma rays from
the location where they are emitted to the outer regions. To calculate
this, we solve the gamma-ray transfer in the gray approximation for any
distribution of $^{56}$Ni,  assuming that gamma rays interact with matter
only through absorption. It has been shown by comparison with Monte Carlo
simulations of Type Ia supernovae that the complex scattering
process between gamma rays and electrons can be satisfactorily 
approximated as an absorptive process
\citep{1984ApJ...280..282S,1995ApJ...446..766S}. The 
value for the gamma-ray opacity that we adopt here is 
$\kappa_\gamma= 0.06 \, y_e$ cm$^{2}$ g$^{-1}$, where $y_e$ is the
number of electrons per baryon. 

\begin{figure}
\begin{center}
\includegraphics[scale=.40]{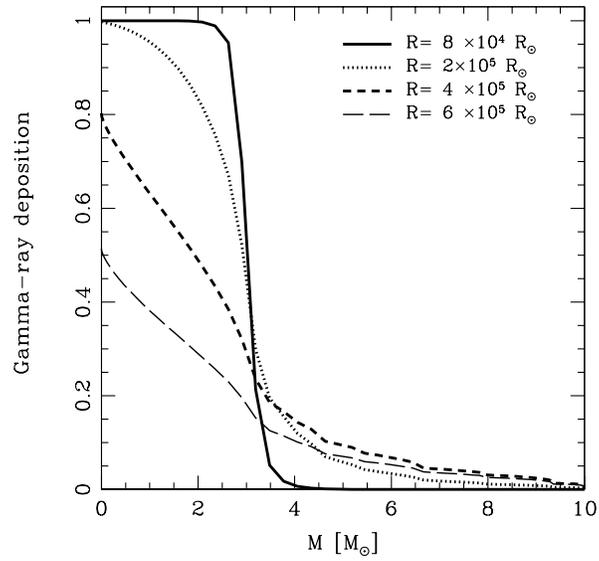}
\caption{Gamma-ray deposition as a function of mass for a
  polytrope with index  n=3, initial mass of 
$10 \, M_\odot$ and different initial radii for a constant
  distribution of $^{56}$Ni up to $3 \, M_\odot$. The 
diffusion of the gamma rays out of the region were they form
becomes more noticeable as the object gets more diluted.
 \label{fig:depo}}
\end{center}
\end{figure}

The rate of energy per gram released by the Ni-Co-Fe decay is 

\begin{equation}
\epsilon_{\rm rad} = \, 3.9 \times 10^{10} \,\, \exp(- t/\tau_{\rm Ni}) +
\, 6.78 \times 10^{9} [\exp(- t/\tau_{\rm Co}) -  \exp(- t/\tau_{\rm
    Ni})] \;{\rm erg\; g}^{-1} \,{\rm s}^{-1}, 
\end{equation}

\noindent where $\tau_{\rm Ni}= 8.8$ days  and $\tau_{\rm Co}=113.6$
days are the mean lifetimes of the radioactive isotopes.
The amount of energy  deposited at each point is given by the
solution of the gamma-ray transfer multiplied by the previous
expression. In Figure~\ref{fig:depo} we show the gamma-ray
 deposition\footnote{The deposition of gamma rays is defined
   by the energy deposited in each point normalized by
   the value corresponding to complete thermalization at the same
   location where the gamma rays are emitted.} for the case
 of a polytrope with index $n=3$, initial mass of
$10 \, M_\odot$ and different initial radii, assuming a constant
 distribution of $^{56}$Ni up to $3 M_\odot$. Note that the diffusion
of gamma rays from the region where they form becomes more noticeable
as the object becomes more extended and diluted. 

The possibility to use an arbitrary distribution of $^{56}$Ni allows
us to study different types of mixing and their effect on the LC. We
have found, as shown in ~\ref{sec:tail}, that
 radioactivity becomes an important source of energy even during
the plateau phase if we allow  extensive  $^{56}$Ni mixing.

\subsection{Initial models}
\label{sec:initialmodel}

There are two different types of initial (or pre-supernova)
models: those coming from stellar evolution calculations
(or ``evolutionary'' models), and those from non-evolutionary calculations
where the initial density and chemical composition are parameterized
in a convenient way. In this work we use double polytropic models in
hydrostatic equilibrium as non-evolutionary pre-supernova
models. Although a single
polytrope may represent very well the envelope of the pre-supernova
object, the inner, dense part which is expected for this type of
objects is not well reproduced. One way to improve this situation is
to consider two polytropes: one representing the inner dense core, and
the other accounting for the outer extended envelope. As initial
composition we use parametric chemical profiles with mixing between
layers of different chemical composition (see Figure~\ref{fig:comp} of
section~\ref{sec:hydro} for SN~1999em).  Mixing is expected
to occur during the explosion due to hydrodynamical instabilities
which can carry the hydrogen very deep into the core and the $^{56}$Ni
out into the hydrogen envelope as previous studies of SN~1987A have
shown \citep[][among
  others]{1987Natur.330..230D,1988A&A...196..141S,1988ApJ...324..466W,1988ApJ...331..377A, 1990ApJ...360..257H,2000ApJ...532.1132B}. Unfortunately,  
in our one-dimensional prescription we 
can not properly take into account this effect. To
ameliorate this we implicitly include this effect by imposing  
 mixing in our initial chemical profile. It is important to
 note that even if mixing is expected only after  
 the shock propagation, the effect on the dynamics of imposing it at
 the initial time is not
noticeable until the shock breakout, when the recombination front
recedes into the ejecta.

In order to calculate numerically a double polytropic model, it is
necessary to solve the equations of hydrostatic equilibrium and mass
conservation assuming a polytropic 
approximation where the  pressure has a dependence on  density
 of the form: $P=K \rho^{\gamma}= K \rho^{n / (n+1) }$, where $K$
is a constant and $n$ is the polytropic index. Two different
polytropic relations with different indices ($n_i$ and $n_o$) and
constants ($K_i$ and $K_o$) are adopted to mimic the characteristic
structure of a red supergiant. To calculate the composite polytrope, the
previous equations are numerically integrated outward from the inner
border (stellar core) to a pre-selected
point (fitting point) using the internal polytropic relation. A
second inward integration is done from the outer border (stellar surface)  
to the fitting point using the external polytropic relation. The
problem is equivalent to a two-point boundary condition for the
case where there are unknown free parameters at both ends of the
domain. In our case, the free parameters are $n_i$, $K_i$ and $K_o$,
while the external index is fixed to $n_o=3$. Starting from
initial guesses, the values of the free parameters are iteratively
sought so that the solution joins smoothly at the fitting point. This
process is performed using a shooting method. Once the values of 
$n_i$, $K_i$ and $K_o$ are found, the density (or pressure) distribution and
mass distribution can be calculated. The initial temperature
profile is calculated in an iterative fashion using our EOS to ensure
hydrostatic equilibrium. This prevents the formation of a spurious
shock wave.  

Using this method a variety of initial models in hydrodynamical equilibrium 
can be calculated by changing the fitting point and the boundary
conditions for a given mass, radius and central density, which allows
us to study how the internal structure of the initial model affects the
LC. 

 Alternatively to the parametric initial model we have tested our code 
using initial models from different stellar evolution
calculations. We found several features in the resulting LC which
are not present in the 
observations, as was previously noted in the literature
\citep[see][]{2008A&A...491..507U}. Specifically, several bumps and
wiggles appear during the plateau phase. Regardless of the
differences among evolutionary models, all of them have the common
characteristic of showing a sharp boundary between layers with
different chemical composition and a steeper jump 
in density between the helium core and the hydrogen envelope than in
our double polytropic models. These two characteristics are the ones
responsible for the unobserved bumps in the LC.  We note that a 
possible reason for the relatively poor results we obtained from evolutionary
initial models may be the use of one-dimensional calculations which
cannot take into account effects that produce mixing, such as Rayleigh-Taylor
instabilities. This point deserves further scrutiny, although it is
beyond the scope of this paper. 

In summary, we decided to use double polytropic initial models based
on the following facts: (1) we have generally found a better agreement with
observations using our parameterized profiles, in accordance to 
previous studies \citep[][among
  others]{1993A&A...270..249U,2005AstL...31..429B,2007A&A...461..233U}; (2)
There is a limited set of pre-supernova models from stellar
evolutionary calculations available to us; (3) Parameterized initial
models allow us to easily vary physical properties and study
  their effect on the LC, which is critical for our goal of
studying a large sample of SNe. However, we emphasize that the choice
of polytropic initial models makes the connection with the structure of a
progenitor star difficult to assess. 

\subsection{Approximations}
\label{sec:assumption}
Several approximations are made in the equations of radiation
hydrodynamics. We assume that the fluid motion can be described by 
one-dimensional, radially symmetric flow. The explosion mechanism of 
core-collapse SNe is not well known but it may be a
very asymmetric process. However, for this particular subtype of
SNe with very extended hydrogen envelopes the asymmetries
expected from the explosion mechanism itself appears to be
smoothed. This is supported by recent spectropolarimetric studies
\citep{2005ASPC..342..330L}. 

We use the equilibrium diffusion approximation to describe the 
radiative transfer. This approximation assumes that radiation and matter
are strongly coupled with a single characteristic temperature and  a 
 spectral energy distribution  described by a black body
function (BB hereafter). The approximation breaks down at shock
breakout and   
at late phases when the ejecta is completely
recombined and the object becomes transparent. Fortunately, during
the plateau phase --- which is our main interest---, this
approximation is adequate. Also note that non-LTE calculations of
SN~II spectra show that deviations from LTE have significant effects
on the lines but not on the overall
continuum \citep{1996MNRAS.283..297B,2008MNRAS.383...57D}. At late phases,
for SNe that experience little interaction with the 
interstellar medium, as is the case of SNe~II-P, the bolometric
luminosity can be simply approximated as the luminosity deposited by
the radioactive 
decay of $^{56}$Co, at least during the early part of the
radioactive tail. This is supported by the fact that the luminosity
declines obeying an exponential law with a very similar rate to that
of the decay of $^{56}$Co. No attempt to apply our model beyond $\sim$180
days is done.

 During  the transition  between  
plateau and radioactive tail, the envelope is fully recombined and
the notion of a photosphere loses meaning. Thus, this transition
regime is poorly described with our radioactive
transfer prescription and therefore a detailed description of this
phase cannot be assessed here. As stated above,
we have performed a comparison of our calculations with those of the
STELLA code, obtaining an excellent agreement
between the bolometric luminosities derived by both
methods. This is very satisfactory considering that the STELLA code
  involves a more sophisticated treatment of the radiative transfer by
  solving these equations using a multi-group prescription and including the
  effect of line opacities \citep{1993A&A...273..106B,2002nuas.conf...57S}. 

\subsection{Comparison with observables}
\label{sec:model_obs}

One of the main motivations for this work is  the unprecedented
database of BVI light curves of $\sim$ 30 SNe which offers the opportunity 
to significantly improve our understanding of SNe~II-P and their
progenitors. Given that our
code produces bolometric
light curves, we need to compute bolometric luminosities for our data set. In
\citet{2009ApJ...701..200B} we found a tight correlation between
bolometric correction and colors which  allowed us to
calculate bolometric luminosities from $BVI$ colors with an 
uncertainty of only $\sim0.05$ dex\footnote{An alternative would be to 
compute $BVI$ light curves  from our models but the  only available 
atmosphere model  that we have at this point is the black body
spectral distribution. However, as several 
studies have shown and as can be seen in Figure~3 of
\citet{2009ApJ...701..200B}, the luminosity of SNe~II-P depart
considerably from a BB behavior very early in the SN evolution, which
hampers the comparison.}.

Another critical parameter to compare with observations is the
photospheric velocity yielded by our models.
Expansion velocity estimates from the minimum of several spectral
lines (H$_\alpha$, Fe {\sc ii} $\lambda5169$, H$_\beta$  and
H$_\gamma$) of our dataset were given by
\citet{2009ApJ...696.1176J}. Since each line forms at a different shell,
it is not straight forward to compare the observed velocities with
photospheric velocities. \citet{2009ApJ...696.1176J} got around this
problem and derived congruent calibrations between
the velocity derived from the absorption minimum of such lines and the
photospheric velocity using  two independent atmosphere models
(\citet{E96} and \citet{D05}; E96 and D05 hereafter, respectively). As noted by
\citet{2009ApJ...696.1176J} H$_\beta$ provides a very good proxy to
the photosphere velocity as it is not highly saturated as H$_\alpha$,
and is present over most of the  SN evolution. Therefore, in
  the comparisons with our models, we chose 
to use the calibration given by \citet{2009ApJ...696.1176J} for this particular
line. However, in spite of the satisfactory behavior of this
  calibration in the high-velocity regime ($v \gtrsim
  5000$ km s$^{-1}$), for lower
  velocities the atmosphere models 
  fail to reproduce the behavior shown by the data \citep[see Figure~9
  of][]{2009ApJ...696.1176J}.
  Thus, we decided to include  also in our comparison the velocities estimated
  from the Fe~{\sc ii} $\lambda5169$ line which is present at later phases of
  the SN evolution when the velocities are below the limit imposed by
  the calibration of hydrogen lines.

 For consistency with the atmosphere models, we define the photospheric
position as the layer where the total continuum optical depth is
$\tau=2/3$. Note that to calculate $\tau$ we do not include the
opacity floor because such minimum is only included to take into account
line effects, and therefore is a bound-bound opacity
that does not contribute to form the continuum. With this
definition, the photosphere follows the recombination wave as we show
in section ~\ref{sec:CRW}. Note that this behavior of the photosphere
is due to the fact that electron scattering is the dominant source of
opacity (see Figure ~\ref{fig:opal}). Had  we
included the opacity floor in the definition of $\tau$,  the
$\tau=2/3$ surface would be pushed well above the recombination front.

We remark that with this definition, the photosphere is essentially 
the surface of last scattering. However, the surface where
the continuum is actually formed is located in a deeper layer called
the ``thermalization depth'' and this is due to the dominance of
the electron scattering over the absorption processes 
\citep{1980Afz....16..695S,1991supe.conf..415H,1995ApJ...445..828M}. With
our simple prescription  of radiative transfer we cannot accurately
determine the 
location of the thermalization depth because there 
radiation decouples from matter. Note that the color
temperature, determined by a black body fit to the broad-band
photometry, is nearly coincident with the temperature of the
thermalization depth but is greater than the effective temperature
defined by $T_{\rm eff}= L / 4 \pi \sigma R_{\rm ph}^2$,  where
$R_{\rm ph}$ is the photospheric radius. In
\citet{2009ApJ...701..200B} we derived a calibration between 
$T_{\rm eff}$  and color based on the E96 and D05 models. Using this
calibration we study the evolution of $T_{\rm eff}$ for our sample of
SNe~II-P and we compare it with  $T_{\rm eff}$  derived from our
code. 

\section{APPLICATION TO SUPERNOVA 1999em}
\label{sec:sn99em}

In preparation for our comparison with
the sample of SNe~II-P in a forthcoming paper, we  
analyze here the case of the prototype SN~1999em 
to show the details of how our  code works. The choice of this object was
motivated by the fact that this is one of the best observed SNe of its
type both in terms of wavelength coverage and temporal sampling. In
addition, SN~1999em was previously modeled with two different 
hydrodynamical codes \citep{2005AstL...31..429B,2007A&A...461..233U} 
which provided consistent physical parameters of the 
  pre-supernova model. Here, we  start our study using our 
code and  initial   
 parameters  consistent with such studies (\S~\ref{sec:hydro}). In
 \S~\ref{sec:compsn99em} we discuss how our results compare with 
 previous studies of SN~1999em.

It is interesting to note  that currently there is an unsolved
discrepancy between the mass of the progenitor object derived from
hydrodynamical models, and  estimates from pre-supernova images
in connection with stellar evolution models
\citep{2008A&A...491..507U,2009MNRAS.395.1409S,2009arXiv0908.0700S}. For
example, for SN~1999em the hydrodynamical mass is close to 19
$M_\odot$ (this corresponds to the mass of the progenitor at the time of
explosion, so the zero-age main sequence star would be even greater) while
from pre-supernova images an upper limit of 15 $M_\odot$ was derived
for the progenitor star at the ZAMS. In order to address this problem,
in \S~\ref{sec:lowmass} we study the possibility of fitting 
the light curve and velocity evolution of SN~1999em with lower values of the
envelope mass.

SN~1999em was discovered  shortly after the explosion on 1999 October 29 UT
by the LOSS program \citep{1999IAUC.7294....1L}. Based on the
Expanding Photosphere Method (EPM) study by
\citet{2009ApJ...696.1176J}, we assume SN~1999em exploded
 3 days before discovery. This is consistent with the
constraint imposed by a negative detection (limiting magnitude 19) on
an image obtained on 1999 October 20 UT. In order to compare with our model,
we corrected the times elapsed since explosion  by time
dilation based on the redshift of the host galaxy. We adopted the  
Cepheid distance of 11.7 Mpc given by \citet{Leo03} to compute bolometric
luminosities as explained in \citet{2009ApJ...701..200B} using
 photometric data obtained at  
{\em Cerro Tololo Inter-American Observatory} (CTIO),  
{\em Las Campanas Observatory} (LCO), and the 
{\em European Southern Observatory} (ESO) at {\em La Silla} \citep{H01}.

\subsection{High-Mass SN~1999em}
\label{sec:hydro}
In this section we use a pre-supernova model with initial parameters
consistent with the optimal hydrodynamical model of
\citet{2007A&A...461..233U}. Specifically, we adopt an initial
mass of 19 $M_\odot$,  radius of  800 $R_{\odot}$, and 
explosion energy of $E=1.25$ foe (1 foe=  $1\times 10^{51}$
erg). This energy was 
released as thermal energy  near  
the core of the object in a very short time scale as compared
with the hydrodynamic time scale of our model. We also assume a
nickel mass of $0.056$ $M_{\odot}$ which was determined by the
luminosity of the radioactive tail. This parameter is quite
different to the nickel mass of $0.036$ $M_{\odot}$ used by
\citet{2007A&A...461..233U} which was determined using the
quasi-bolometric luminosity given by \citet{2003A&A...404.1077E}. Their
luminosity was based on the integration of only $UBVRI$ photometry with
a constant value of $0.19$ dex added to take into account the infrared
luminosity, while our bolometric luminosity is based on a quantitative study
of the bolometric correction for SNe~II-P (see
\S~\ref{sec:assumption}). Note also that our value for the nickel mass
is closer to the estimate of $0.06$ $M_{\odot}$
given by \citet{2005AstL...31..429B}. In the following analysis, we
denote this model as \mod.\\

\begin{figure}
\begin{center}
\includegraphics[angle=-90,scale=.40]{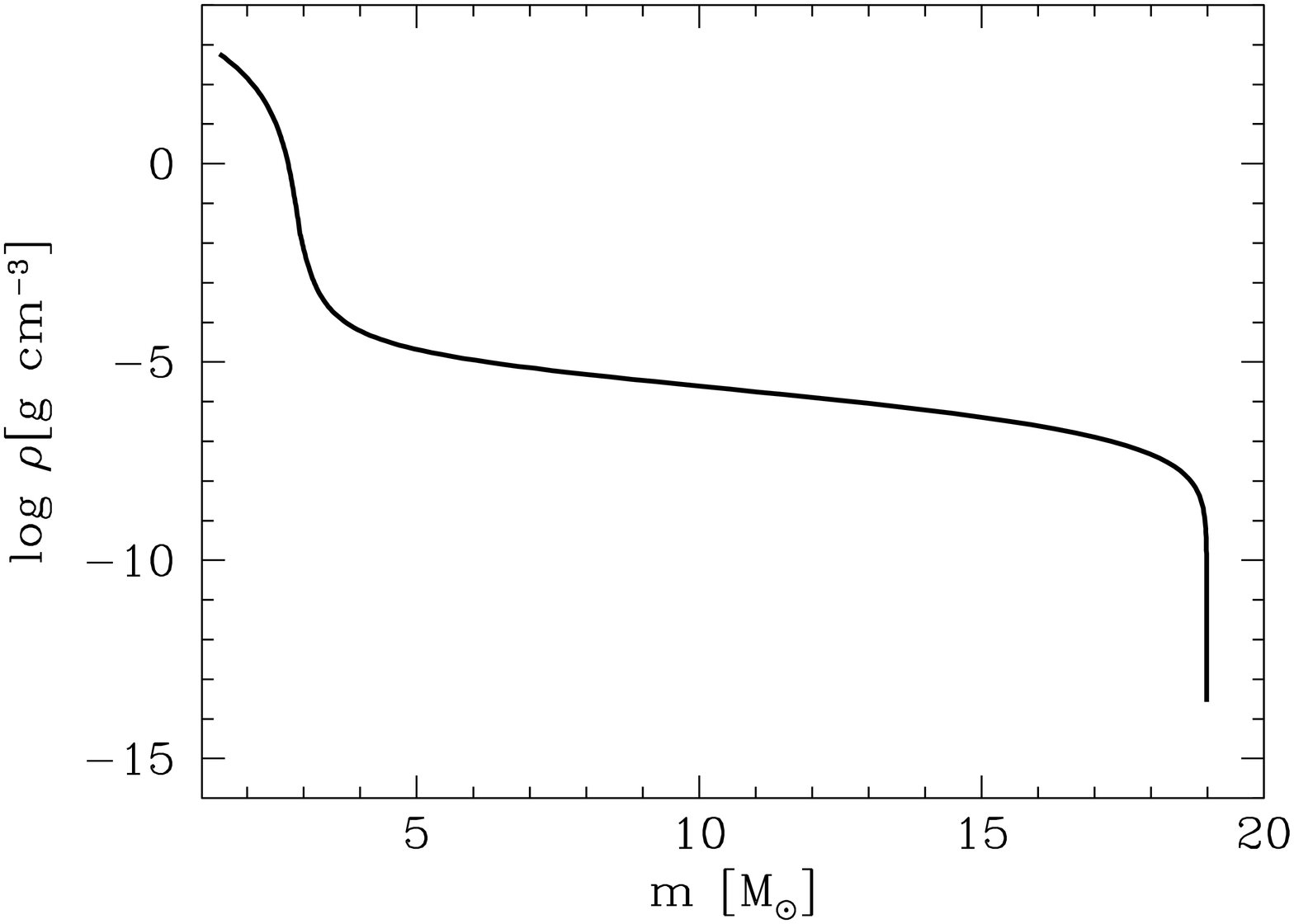}
\includegraphics[angle=-90,scale=.40]{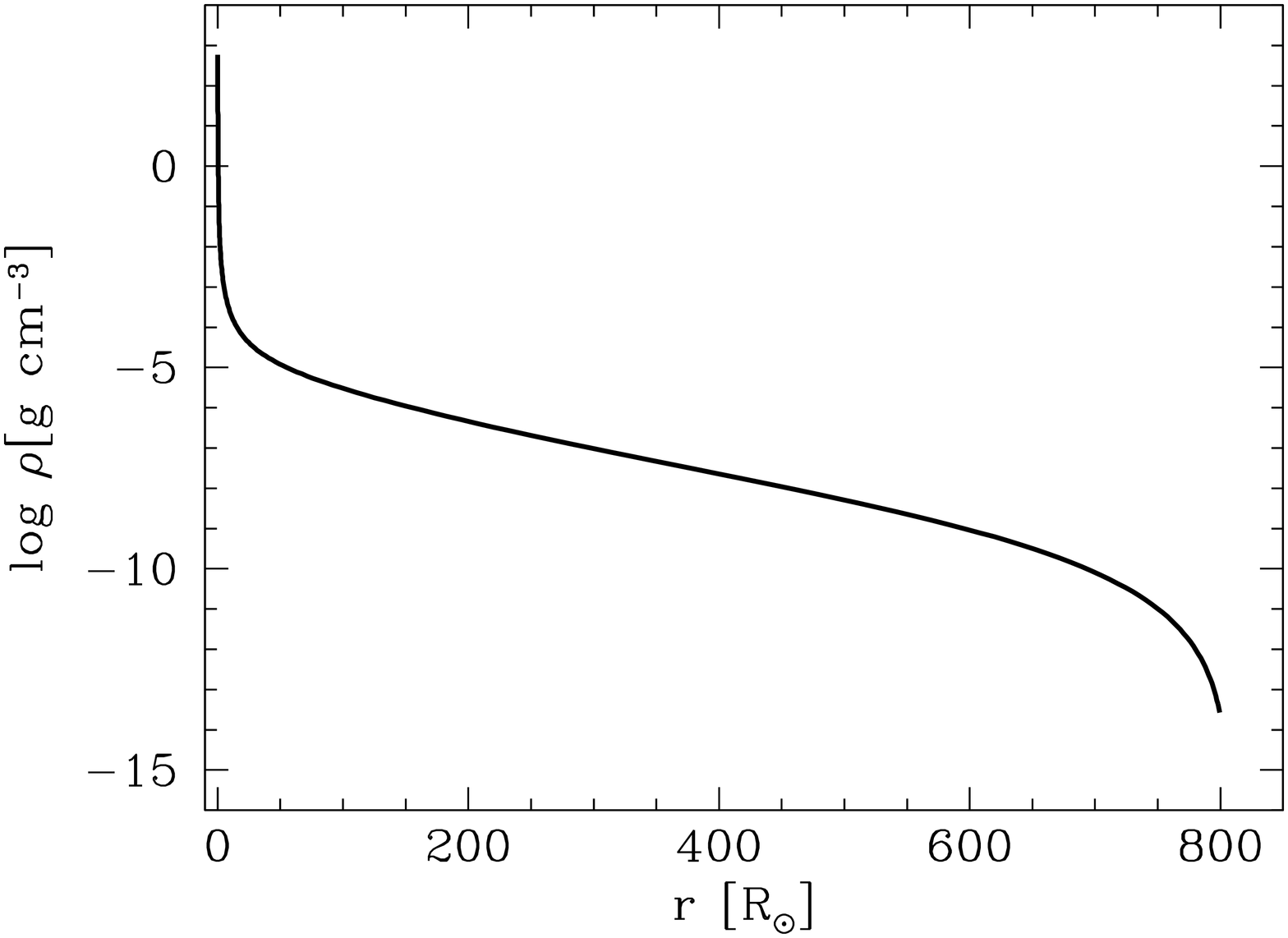} 
\caption{Initial density distribution with respect
  to interior mass ({\em top}) and radius ({\em bottom}) for
  the pre-supernova model \mod.\label{fig:dens}}
\end{center}
\end{figure}

\begin{figure}
\begin{center}
\includegraphics[angle=-90,scale=.40]{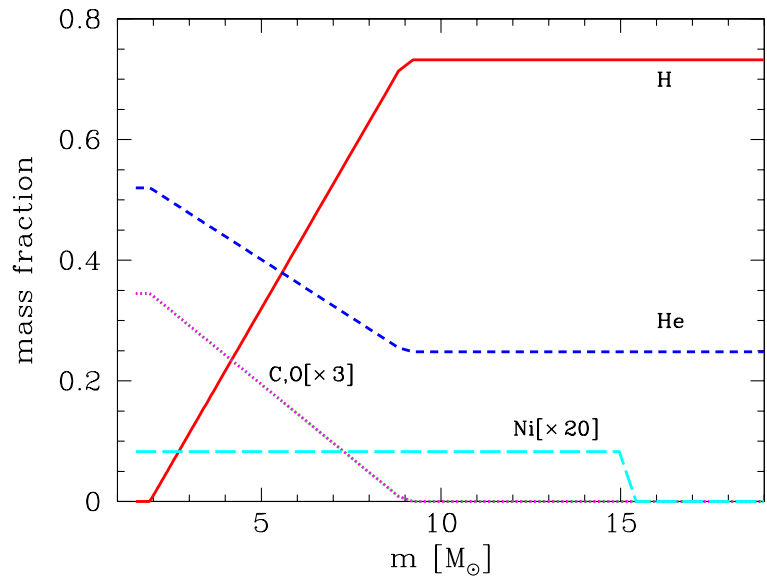}
\caption{Abundance distribution in the pre-supernova model
  \mod\ with respect
  to Lagrangian mass. For clarity, the abundance of
  carbon (C) and  oxygen (O) were multiplied by 3 and the abundance of
  $^{56}$Ni, by 20. Note that the $^{56}$Ni is uniformly mixed in the
  outer envelope until 15 $M_\odot$. \label{fig:comp}} 
\end{center}
\end{figure}

In our calculations we remove the central $1.4$ $M_\odot$ which is assumed to
form a neutron star. The initial density profile as function of mass
and radius for model \mod \  is shown in
Figure~\ref{fig:dens}. Note that the initial 
structure is composed of a dense core
and an extended envelope characteristic of a red supergiant. The
chemical composition profiles are shown in Figure~\ref{fig:comp}. The
outer parts of the envelope, $M > 9 \, M_\odot$, 
have a homogeneous composition with a mass fractions of
$X=0.735$,  $Y=0.251$, and 
$Z=0.02$. From there inward, hydrogen and helium are mixed in order to
prevent a sharp boundary between the H-rich and
the He-rich layers. Such sharp boundaries are characteristic of
stellar evolution models but fail to reproduce the observations. Note
that we allow H to mix very deep inside the core and  $^{56}$Ni is mixed 
out in the envelope until $\sim$15 $M_\odot$. This type of mixing is
presumably due to Rayleigh-Taylor
instabilities that occur behind the shock front, as is obtained in
multi-dimensional hydrodynamic calculations
\citep{1991A&A...251..505M,2000ApJ...528..989K}
and supported  
 by studies of SN~1987A \citep[][among others]
{1988A&A...196..141S,1988ApJ...324..466W,1988ApJ...331..377A,2000ApJ...532.1132B}. The
presence of H in the core   
leads to a smooth transition between the plateau and the radioactive
tail. The distribution of $^{56}$Ni to external layers helps to
reproduce a plateau as flat as that observed in SN~1999em (see
section~\ref{sec:CRW}). 

\begin{figure}
\begin{center}
\includegraphics[angle=-90,scale=.40]{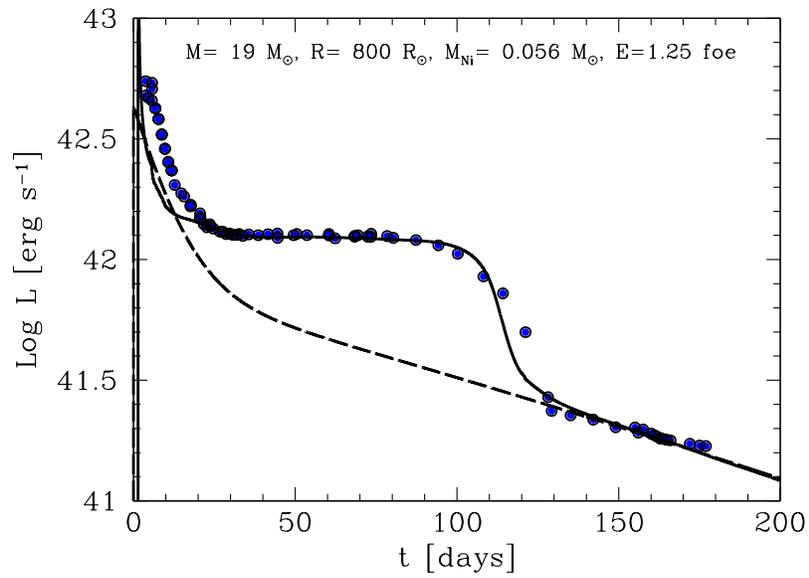}
\caption{ The bolometric light curve for  model 
 \mod\ (solid line) compared with the data of SN~1999em as calculated
 by \citet{2009ApJ...701..200B} (blue dots). The 
  luminosity due to the $^{56}$Ni  $\rightarrow$ $^{56}$Co
  $\rightarrow$  $^{56}$Fe decay is  
 also shown (dashed line). The physical parameters used in the model  
 are indicated. \label{fig:LC}} 
\end{center}
\end{figure}

Figure~\ref{fig:LC} shows a comparison between the bolometric light
curve (solid line) obtained with our code for model
\mod, and the observations of SN~1999em (dots). The luminosity
due to $^{56}$Ni $\rightarrow$ $^{56}$Co $\rightarrow$ $^{56}$Fe
(dashed line) is also shown. Note the very good 
agreement between model and observations. The largest differences
appear during the earliest phase and  the transition to the
radioactive tail. At the earliest epochs our bolometric corrections have
the largest uncertainties because during this time the UV flux
  ---which is the main contribution of the luminosity--- is not well
  constrained as noted in \citet{2009ApJ...701..200B}. During the transition 
between the plateau and the radioactive tail, the
diffusion approximation breaks down because
the object is almost completely recombined and the photosphere is not well
defined. During the tail, the bolometric 
luminosity is completely determined by the luminosity of radioactive decay,
which makes our calculations more reliable. Although not shown here,
our model shows that the shape  
of the light curve at the end of the plateau is very sensitive to the
properties of the core, such as mixing and the form of the 
density transition between the helium-rich core and the hydrogen-rich
envelope. On the other hand, the
presence of an outer atmosphere can affect the shape of the
light curve (LC) at the earliest stages. During these epochs, and until the
hydrogen recombination sets in, the LC samples the outermost layers
as the photosphere is located far out in the object (see
\S~\ref{sec:CRW}). So, even it was possible to find a better 
fit to the observations by modifying the innermost or
  outermost structures, such efforts are rendered meaningless due to
  the uncertainties introduced by our approximations.

\begin{figure}
\begin{center}
\includegraphics[angle=-90,scale=.40]{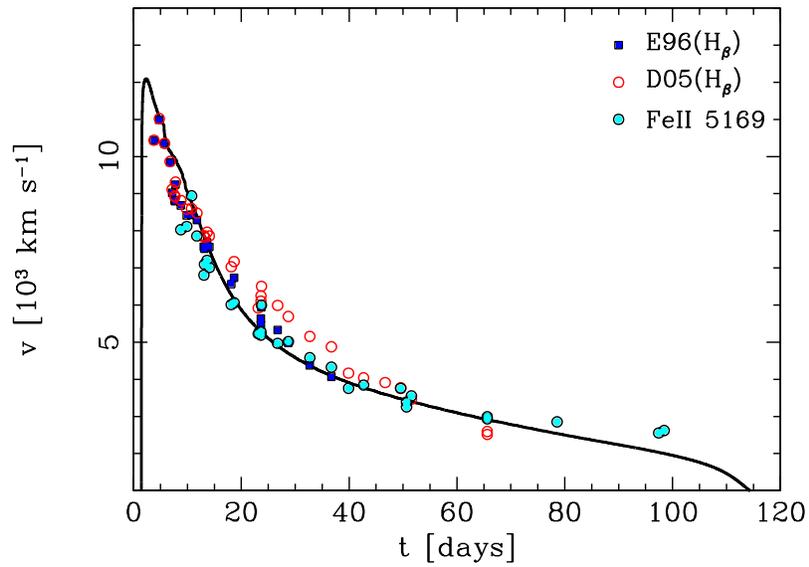}
\caption{Evolution of the expansion velocity of the photosphere
for model \mod\ (solid line) compared with observed  photospheric
velocities  calculated by \citet{2009ApJ...696.1176J} using a
calibration between the velocity derived from the absorption minimum
of H$_\beta$  and two atmosphere models: E96 (filled squares) and D05
(open circles). We also include the velocities calculated from 
Fe~{\sc ii} $\lambda5169$ line (filled circles) which are not transformed to
photospheric velocity. \label{fig:velo}}
\end{center}
\end{figure}

Figure~\ref{fig:velo} shows the photospheric-velocity evolution of our
model compared with observed photospheric velocities as explained in
\S~\ref{sec:assumption}. We have also included in the plot the  
spectroscopic velocities measured from the absorption minimum of
the Fe~{\sc ii} $\lambda5169$ line \citep{2009ApJ...696.1176J}.
Note the very good overall agreement between model
and observations. 
Fe~{\sc ii} velocities match quite well the model photospheric
velocities except at the latest times. However, it is
important to remark that the photospheric velocity is to
be considered a good discriminator between models only at the early
phases of the evolution while the object is not completely recombined. This  
is because, line velocities become poor photospheric
velocity indicators with time, and the photosphere 
begins to lose its meaning in our models as the ejecta becomes nearly
completely recombined at the end of the plateau.

As shown in Figures~\ref{fig:LC} and \ref{fig:velo} we
obtain a remarkably good agreement with observations (bolometric
light curve and  photospheric velocity evolution) for SN~1999em
despite the simplifications used in our code. Note that in spite
  of the mismatch between model and observations during the early LC, the
  photospheric velocities, which are better determined than the LC,
  show a good agreement at such epochs. These results are very
encouraging and give us confidence in the hability of our code to
infer physical parameters, to study their effect on the observed
quantities, and ultimately to understand  the physics of SNe~II-P.

In the following subsections, we describe in some detail
the evolution of the SN for this particular model. At a glance, the LC is
distinguished by three phases: a) an outburst followed by strong
cooling, b) a plateau, and c) a radioactive tail. Each phase is
essentially determined by the interplay between the main 
heating and cooling mechanisms. We thus focus our discussion on
these processes along the SN evolution. As a reference,
Table~\ref{tbl-1} gives a summary of some
properties of the model at characteristic times during the
evolution. Even though some uncertainties may arise on the detailed
propagation of the shock wave due to the assumed mixing and energy
injection, and also on the shock breakout because of the adopted
radiative transport, we think that it is instructive to provide a
description of the processes which occur during such phases. We
emphasize that the details of the shock propagation do not directly
affect the resulting LC and photospheric velocities, which we are 
mainly interested in modeling.

The initial phase of shock breakout is discussed in \S~\ref{sec:SW}, the
adiabatic-cooling phase where the homologous expansion is reached is
addressed in \S~\ref{sec:AC}, the cooling and recombination phase is
described in \S ~\ref{sec:CRW}, and finally the radioactive-decay
processes that power the tail are explained in \S~\ref{sec:tail}.

\subsubsection{Shock wave propagation and the early evolution}
\label{sec:SW}
A powerful shock wave begins to propagate outward through the envelope when
we artificially inject energy near the center of the star (assumed to occur at
$t=0$). This energy is initially released as internal energy, and
part of it is rapidly transformed into kinetic energy (e.g., by $t=0.5$~days,
the total energy is  approximately equally divided between kinetic
and internal energy). The velocities acquired by matter are so high
that they exceed the local speed of sound, leading to the formation of
a shock wave. The shock wave heats and accelerates the matter
depositing mechanical 
and thermal energy into successive layers of the envelope until it
reaches the surface, where photon diffusion dominates the energy
transfer, and energy begins to be radiated away. 

\begin{figure}
\begin{center}
\includegraphics[angle=-90,scale=.30]{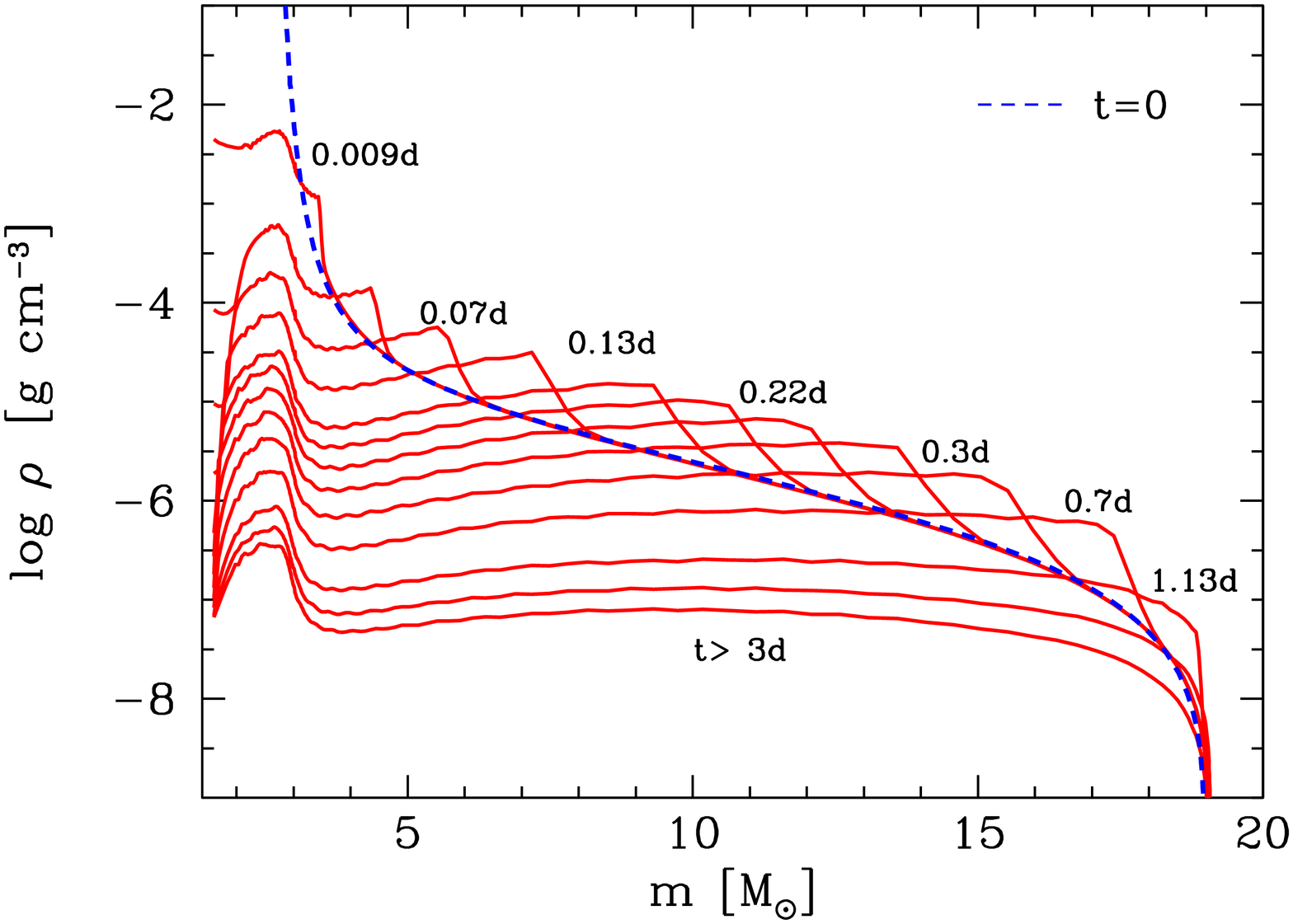}
\includegraphics[angle=-90,scale=.30]{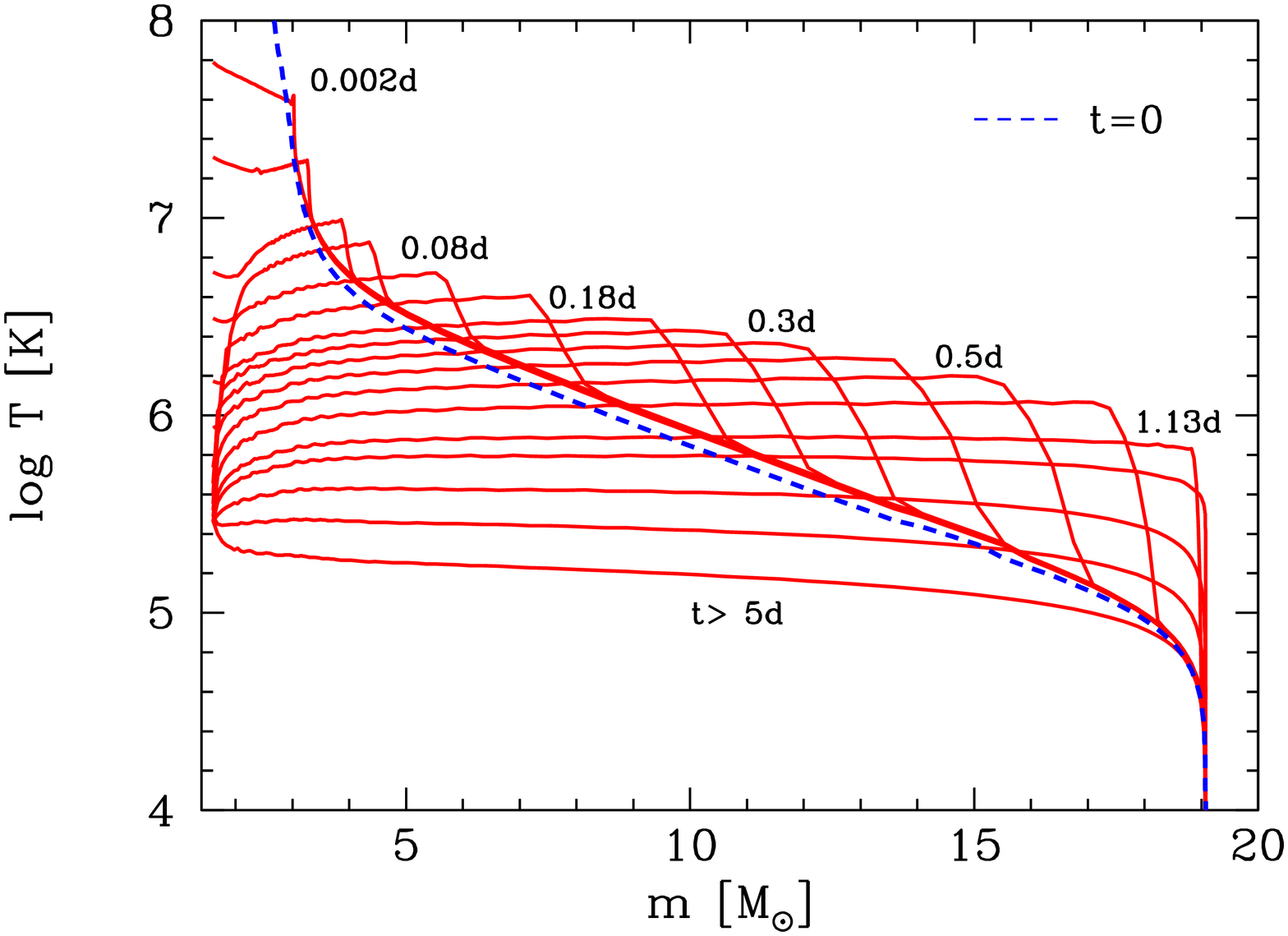} \\ 
\includegraphics[angle=-90,scale=.30]{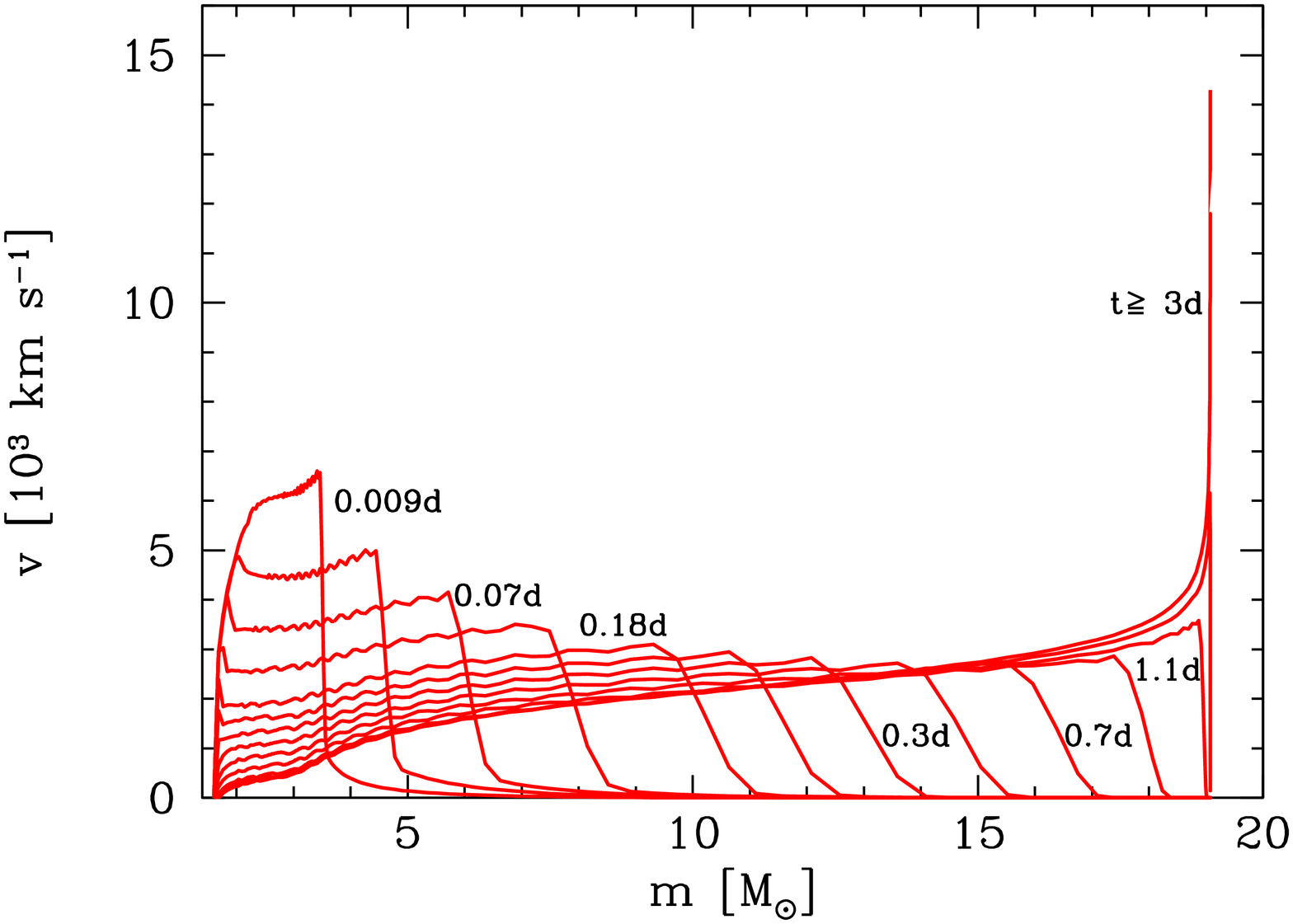}
\includegraphics[angle=-90,scale=.30]{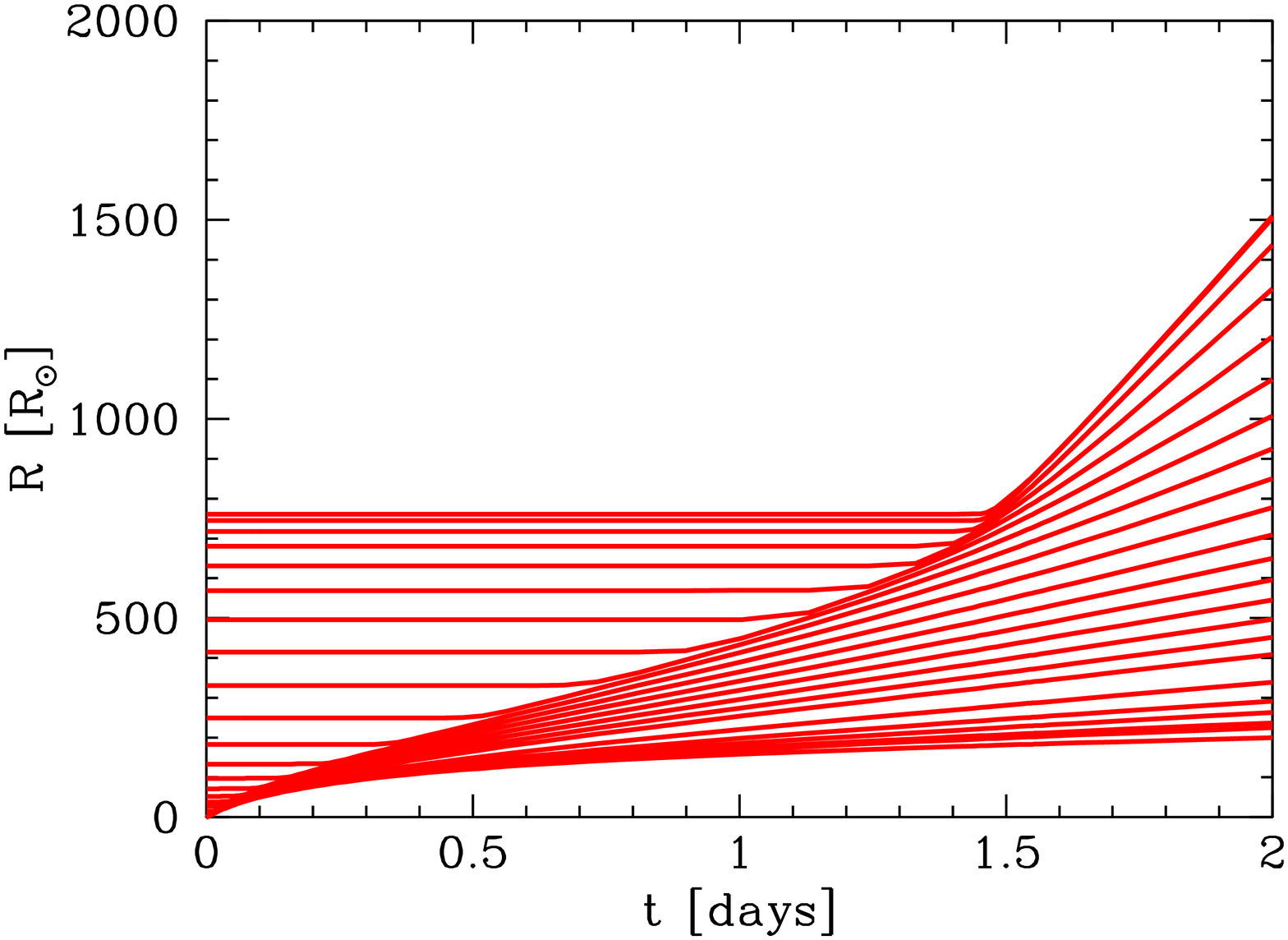} 
\caption{ Changes in  density {\bf  (top left)},
  temperature {\bf  (top right)}, and velocity {\bf  (bottom left)}
  profiles  as a function of interior mass during  shock
  propagation for model \mod. Some of the curves are labeled with the
  time elapsed since the energy is injected. The initial density and
  temperature  profiles are also shown ($t=0$; dashed line). Note that a very
  small amount of material near the surface is strongly accelerated as
  the shock wave passes through the steep density gradients present in the
  outermost layers.   
  {\bf Bottom right:} temporal evolution of
  the radial coordinates corresponding to different mass shells in the
  interior. \label{fig:profiles}}
\end{center}
\end{figure}

Figure~\ref{fig:profiles} shows the effect of the shock wave
propagation on different 
physical quantities (velocity, density and temperature) inside the star. The
shock front  manifests as a sudden change in these quantities. As the shock
moves outward, the material behind is accelerated and
heated. Note that the outermost layer, with a sharp decline in density, 
acquires very high velocities. It represents a small fraction of
the star. 
It is also clear from the velocity profiles that the inner
parts of the object are decelerated. This deceleration is due to the
interaction of the dense core with the extended hydrogen-rich
envelope. The same figure also shows the changes in radius for
different shells. From this we can deduce the location of the shock
front at any time as that of the innermost shell which has constant
radius. 
                
At $t=1.36$~days, the shock wave reaches the surface of the object, which
produces the first electromagnetic manifestation of the
explosion (although neutrinos and gravitational waves escape well
before). The effective temperature and bolometric luminosity  suddenly
rise and reach their maximum values a 
few hours after  breakout, specifically, at $t=1.47$~days with
values of $L_{\rm peak}= 3\times 10^{44}$~erg~s$^{-1}$ and
$T_{\rm peak}=1.1\times 10^{5}$~K.
At these  temperatures, the peak of the emitted  spectrum
is in the UV range. Also note than the color
temperature is even higher than the effective temperature, as mentioned
in section~\ref{sec:assumption}.

\begin{figure}
\begin{center}
\includegraphics[angle=-90,scale=.3]{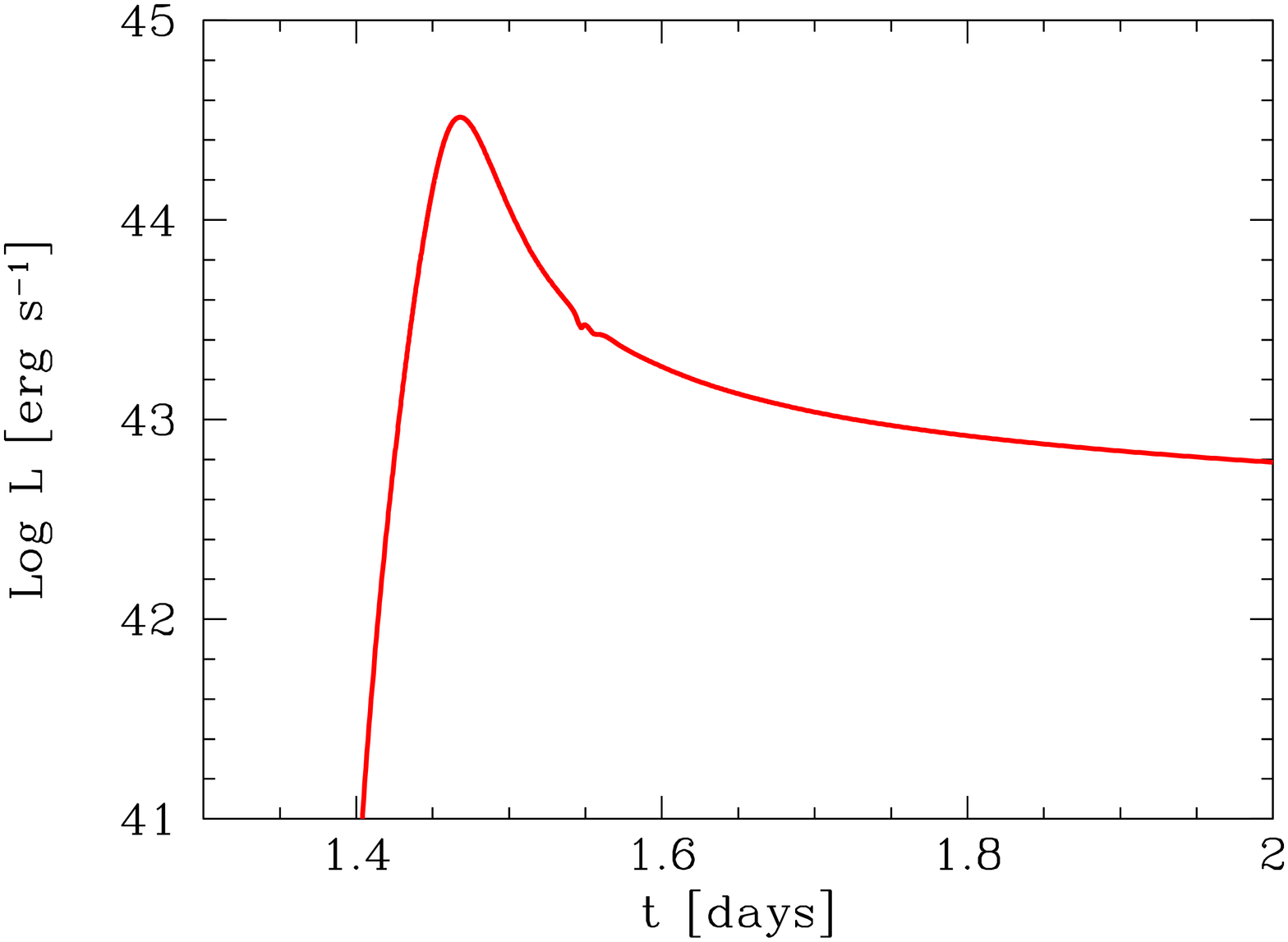}
\includegraphics[angle=-90,scale=.3]{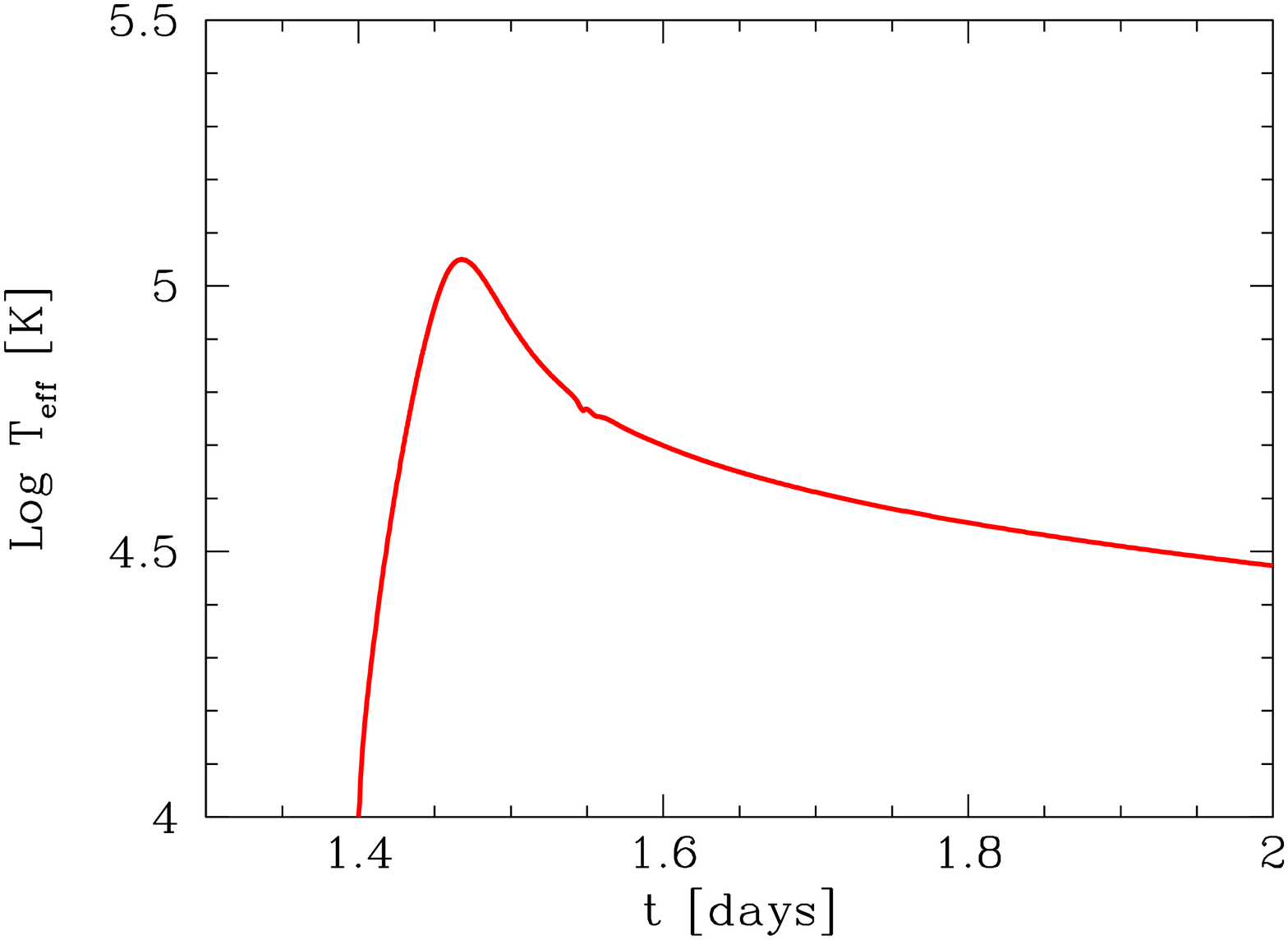}
\caption{Bolometric luminosity ({\bf left}) and effective temperature 
  ({\bf right}) during shock breakout \label{fig:LTinc}}

\end{center}
\end{figure}

Hereafter, the star begins to expand and cool very
quickly, leading to an increase in photospheric radius and a
decrease in temperature in the external layers. The bolometric
luminosity abruptly decreases but, according to the decrease in
effective temperature and consequent shift of the emission peak
to longer wavelengths, the luminosity in the optical range 
increases. As a result, a sharp peak in bolometric luminosity and
temperature is produced, as shown in Figure ~\ref{fig:LTinc}. For
example in our model we obtain a 
  decrease of $1.5$ dex in luminosity only $6.8$ hours after 
  peak brightness.  During this time the total energy radiated is $2.2\times
10^{48}$~erg, emitted essentially as a UV flash. The short
duration of the breakout explains why so few SNe~II-P have been
observed during  this phase. 

At temperatures as high as those left by the passage of the
  shock wave, the stellar 
matter is completely ionized  which implies that  the
breakout is accompanied by 
a strong increase in opacity. The position of the photosphere
during the outburst nearly coincides with the outermost shell.
This behavior continues until the onset of recombination. Therefore,
at early stages previous to recombination, the velocity 
of matter in the photospheric position samples the very high velocities of the
outermost layers and reaches values close to $1.2\times
10^4$~km~s$^{-1}$.

 From the energetics  point of view, by $t=0.5$~day the total energy is
approximately divided in equal proportions between internal, dominated
by radiative contribution, 
and kinetic components. Short after breakout the kinetic energy
completely dominates the energetics (the energy radiated away   
at any given time  is less than $1.5$\% of the  total energy).

 It is possible to estimate the average velocity during the shock wave
propagation: given that the shock wave takes $1.36$~days to emerge and
considering a 
radius of $R= 5.5\times 10^{13}$~cm (800~$R_{\odot}$), we obtain an average
speed of $v=4680$~km~s$^{-1}$. It is also
 interesting to calculate the expected  time for breakout using the
 analytic expression given by \citet{1987Natur.328..320S}, 

\begin{equation}
t_{\rm bk} \simeq 1.6 \left( \frac{R_0}{50 R_\odot} \right)  
\times \left[ \left(  \frac{M_{\rm ej}}{10 M_\odot} \right) / \left(\frac{E}{1
    \times 10^{51} \,\,{\rm erg}} \right) \right]^{1/2}  \,\, {\rm hr},
\end{equation}

\noindent where $R_0$ is the initial radius, $M_{\rm ej}$ is the
ejected mass, and $E$ is the explosion energy. Using the values for
our initial model we obtain $t_{\rm bk}=1.26$~days,  in 
good agreement with our numerical calculation. 

\subsubsection{Adiabatic cooling and homologous expansion}  
\label{sec:AC}
The breakout is followed by a violent expansion, resulting in the cooling of
the outermost layers. During expansion, only a small fraction of the photon
energy can diffuse into the surroundings. Therefore, it is possible to consider
the cooling process to be approximately adiabatic and this approximation remains
valid while the time-scale for radiation diffusion is much longer than the
expansion time-scale\footnote{The expansion time-scale, $\tau_{\rm h}=
R/v$, increases with time while the diffusion time-scale, $\tau_{\rm d}=
\kappa \rho R^2 /c$, decreases because $\rho \propto R^{-3}$. Thus,
there is a time after which the condition for  this approximation
breaks down.}. Note that there are two mechanisms to cool a SN: 
(a) loss of photons or diffusion cooling, and (b) its own expansion or
``adiabatic cooling''. If one of these processes  dominates  we
say that the cooling is 
carried out by that process. In our model, more than 90\% of the
decrease of internal energy is due to adiabatic cooling up to 18 days
after the explosion, when the first layers with 
neutral hydrogen appear. After that, diffusion cooling begins to be
significant and the adiabatic cooling phase comes to an
end. 

The internal temperature (and internal energy) decreases almost
adiabatically, i.e.  proportional to $r^{-1}$ due to the dominance of
the radiative term, and quickly reaches a value near 
the  recombination temperature of hydrogen. Quantitatively, the effective
temperature goes from values close to $10^{5}$~K at the time of the
burst when the matter is totally ionized, to  $T_{\rm eff}=
10^{4}$~K five days later. At these temperatures hydrogen begins
to recombine. At the same time, the luminosity reaches 
$L= 1.9 \times 10^{42}$~erg~s$^{-1}$ and the photospheric radius
rapidly increases to $R_{\rm ph}= 5.2 \times
10^{14}$~cm. After that, the luminosity decreases slightly as a result
of the slower decrease in temperature and continuous increase in
radius. By day 18, when the first layers of neutral hydrogen appear,
we have $T_{\rm eff}=6960$~K, $L= 1.2\times 10^{42}$~erg~s$^{-1}$, and
$R_{ph}= 9.1 \times 10^{14}$~cm. Note the slight change in luminosity,
of only $0.14$ dex, between day 7 and 18.

\begin{figure}
\begin{center}
\includegraphics[angle=-90,scale=.40 ]{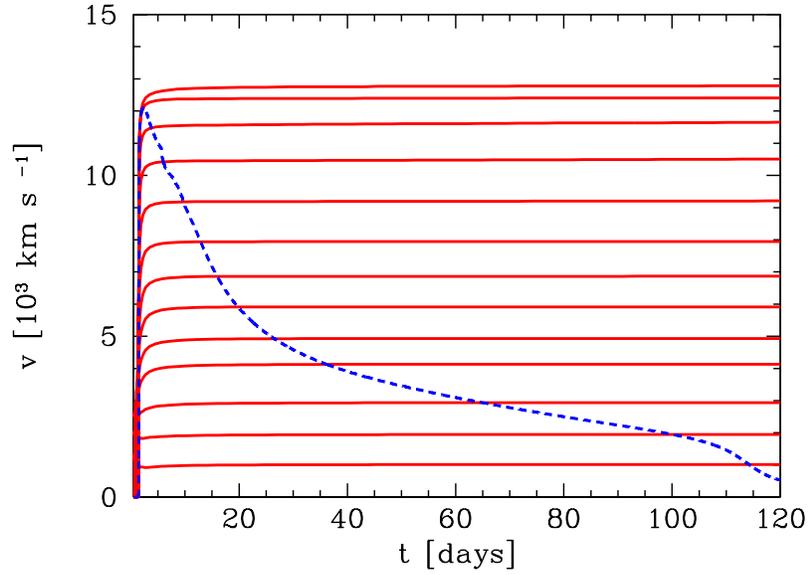}
\caption{Evolution of the velocity for different layers
  (solid lines) and the velocity of material at the photospheric position
  (dashed line). Note that the asymptotic constant velocity is reached at
  a different time for each shell.\label{homovelo}}
\end{center}
\end{figure}

A few days after breakout, the acceleration of the material comes to
an end and the expansion becomes homologous. The
matter reaches velocities of the order of $1.2 \times 10^{4}$~km~s$^{-1}$ in the
outermost layers and the  object enters a state of free expansion where
forces of pressure and gravitation do not have any dynamical effect on the
system. The homologous regime is
characterized by a constant velocity in each layer, a linear growth of
the radial coordinate with time ($r \propto t$) and a
density distribution decreasing with time as $\rho \propto t^{-3}$.  This
behavior is clearly shown in Figure~\ref{fig:profiles} for $t>3$
days and in Figure~\ref{homovelo}. The
nearly constant shape of the density and temperature profiles at 
late times are a result of the expansion being approximately
homologous. This is also evident in the linear behavior of the radial
coordinates for different mass shells. The condition
of constant velocity is very nearly satisfied for each
shell, although the transition between acceleration and homologous
expansion happens at slightly different times for different mass shells
(see Figure~\ref{homovelo})\footnote{After day 8, the
    photosphere remains located in shells where homology has been reached.
   Therefore, our models show that the EPM, which is used to estimate
  distances assuming homologous expansion, must be  applied  at epochs
  later than this.}. 

As  mentioned  in \S~\ref{sec:SW}, at the time of breakout the
internal energy of the envelope is small compared to the kinetic
energy. Moreover, the object expands by a factor of $\sim$ 17 to a
radius of $R_{\rm ph} \sim 9 \times 10 ^{14}$~cm,
before it becomes transparent. Thus,
most of the internal energy left by the 
shock passage is degraded by the adiabatic expansion, and only a small
residual will be released as observable radiation in the following phase. 

\subsubsection{Cooling and recombination wave}
\label{sec:CRW}

\begin{figure}
\begin{center}
\includegraphics[angle=-90,scale=.30]{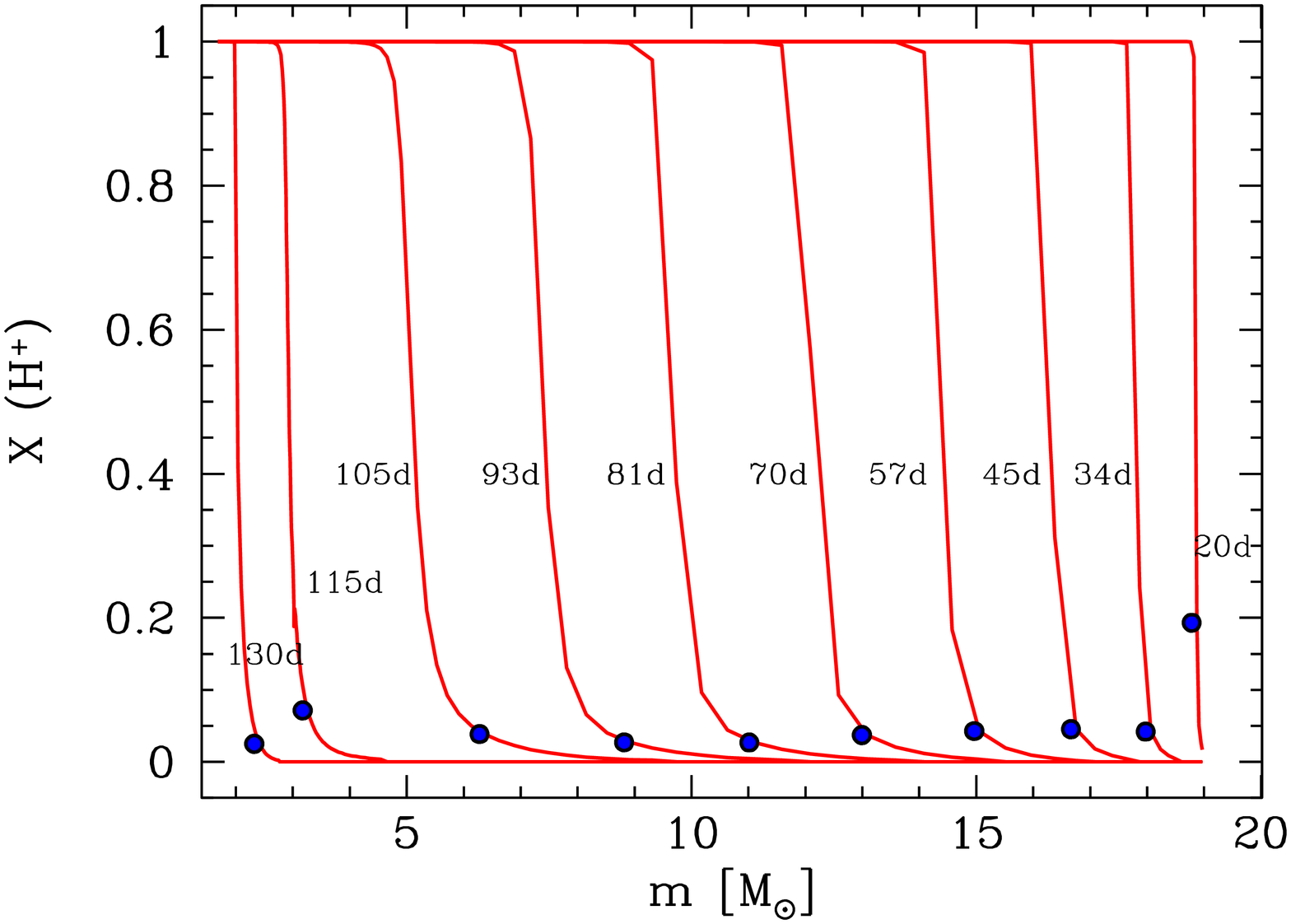}
\includegraphics[angle=-90,scale=.30]{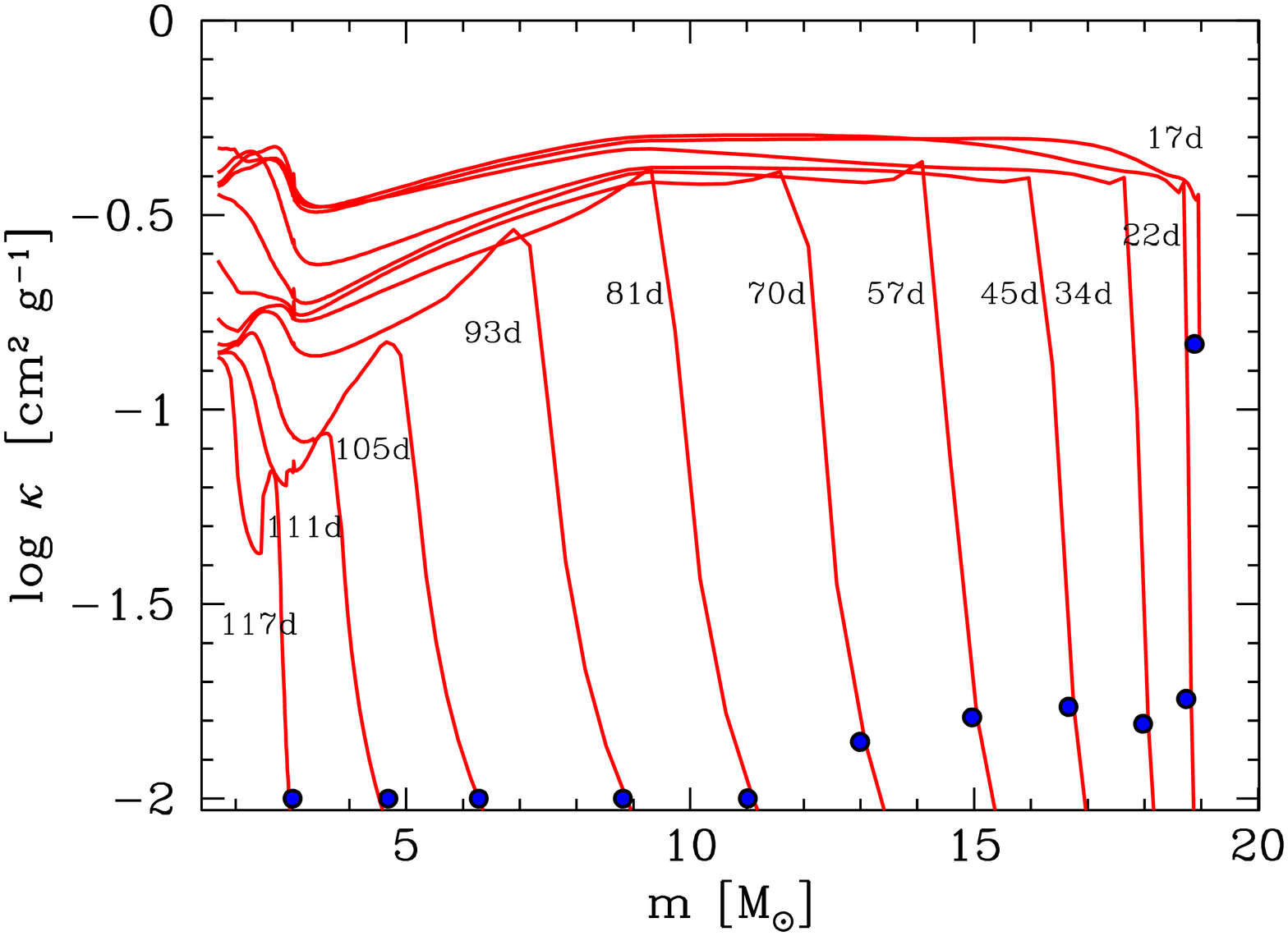} \\ 
\includegraphics[angle=-90,scale=.30]{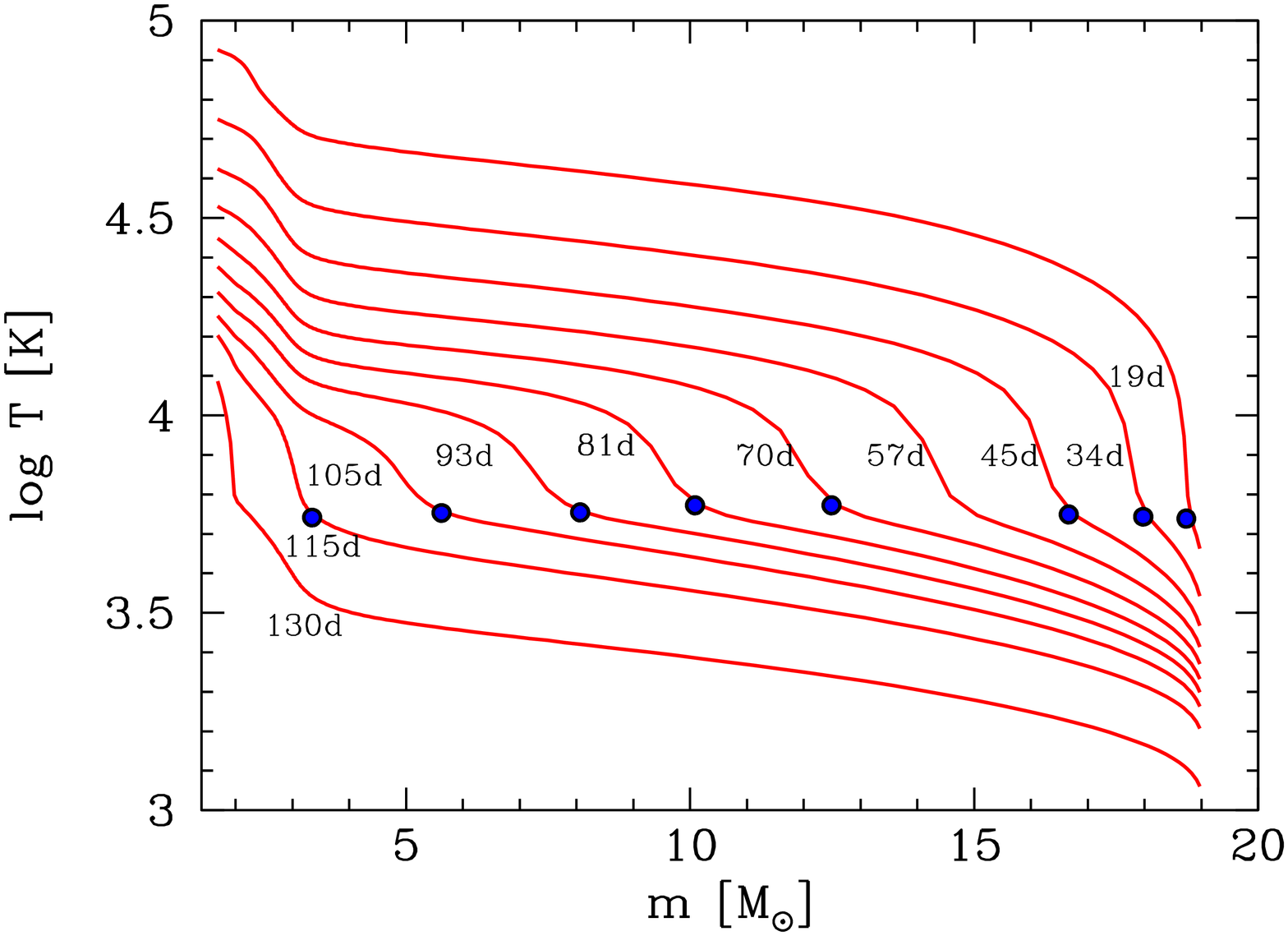}
\caption{Evolution of the fraction of ionized hydrogen {\bf(top left)},
 opacity {\bf  (top right)}, and temperature {\bf(bottom left)}
  as a function of mass. The time elapsed since the energy
  is injected and the photospheric position (blue dot) are indicated
  for each curve. The photosphere  
  is nearly coincident with the outer edge of the CRW which moves
  inward in mass coordinate. Note also that the opacity in the outer
  parts is nearly constant with a value of 0.4 cm$^{2}$ g$^{-1}$,
  close to the electron-scattering opacity for matter
  composed by pure hydrogen, above which it suddenly
  drops.\label{fig:profilesXOT}}  
\end{center}
\end{figure}

The appearance of regions with neutral hydrogen 
determines the onset of a recombination wave (RW). The
duration of the RW is related to the total hydrogen mass and how
deep hydrogen has been mixed in the  initial model.  
As time goes on, hydrogen recombination occurs at different layers
of the object as a wave propagates inward (in Lagrangian coordinate; see
Figure~\ref{fig:profilesXOT}, top left). Because the opacity is dominated by
electron scattering, it strongly decreases outward  the
recombination front (Figure~\ref{fig:profilesXOT}, top right), increasing the
transparency of these layers and allowing the radiation to easily go
away. Consequently, the internal energy (mostly of the radiation field) is
efficiently radiated away, and the temperature drops sharply 
from $\sim$$10000$~K to $\sim$$5500$~K at the recombination front 
(Figure~\ref{fig:profilesXOT}, bottom left). In other words, a cooling wave
associated with the transparency and induced by a recombination wave propagates
through the envelope. This is usually called ``cooling and recombination wave''
(CRW). 

In our model the recombination begins approximately at day
  18. After that, a
CRW develops which moves inward in mass until all the matter is completely
recombined by day 120 when the recombination front arrives at the
innermost layers of the H-rich envelope (at $m=$2 $M_\odot$ for our
model), and the luminosity suddenly drops, defining the end of this
phase. The propagation of the CRW is clear from a glance at
Figure~\ref{fig:profilesXOT} where the evolution of 
the fraction of ionized hydrogen and temperature profiles as a function of mass
for selected times are shown. Note also the behavior of the opacity (see
Figure~\ref{fig:profilesXOT}, top right) which depends strongly on the
ionization state of the matter. In our model, a strong drop in the
opacity of the outer layers is seen at day $\sim$18, which 
leads to a considerable decrease in the 
optical depth of these layers and  an inward drift in mass of the
photosphere --- defined at a fixed optical depth of $\tau = 2/3$.
The photosphere begins to follow the CRW, as shown in Figure
~\ref{fig:profilesXOT} where 
the points indicate the photospheric position. It is important to note 
that we define the position of the photosphere using only continuum
opacity sources, i.e. excluding the opacity floor (see
section~\ref{sec:assumption}).

\begin{figure}
\begin{center}
\includegraphics[angle=-90,scale=.30]{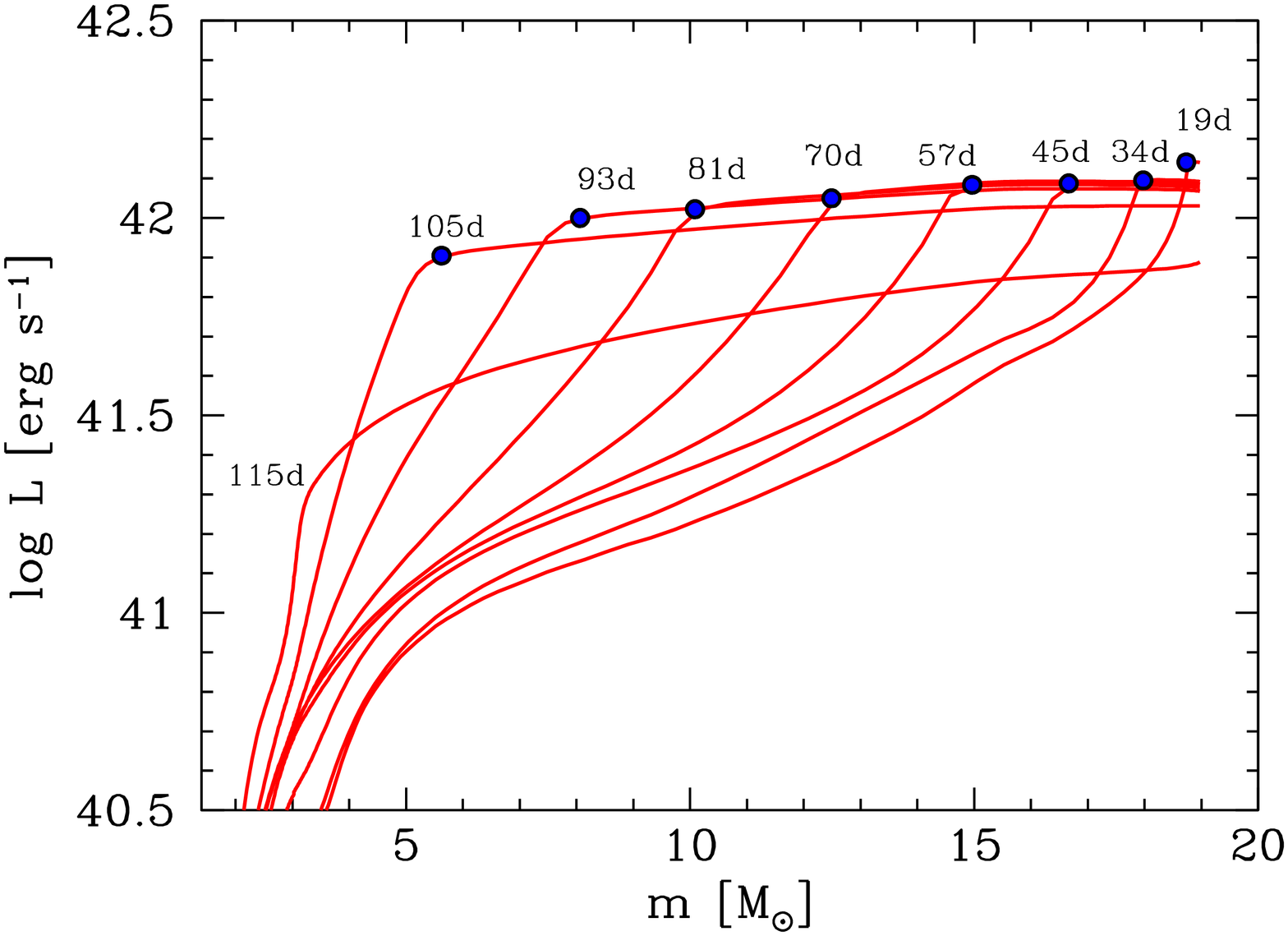}
\includegraphics[angle=-90,scale=.30]{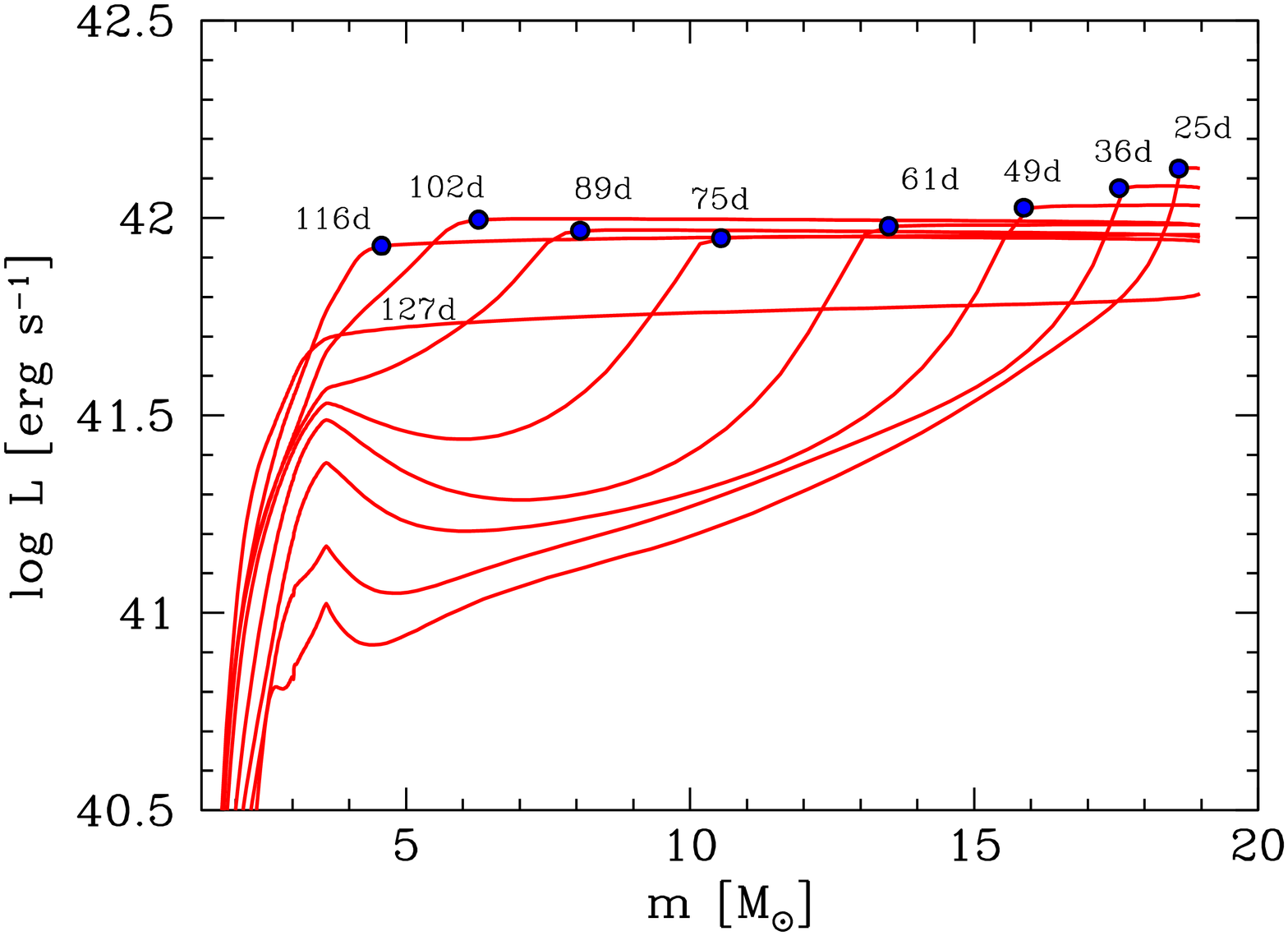} \\ 
\caption{Evolution of the interior luminosity as a function of
  mass. The  labels indicate
  the time since explosion. {\bf Left:} $^{56}$Ni is mixed into the
  hydrogen-rich envelope up to 15 $M_\odot$. {\bf Right:} $^{56}$Ni is
  confined to the   
   layers inside 3.5 $M_\odot$. The inward propagation
   of the recombination front is clear in both mixed and unmixed
   cases. In the mixed case, the recombination front propagates more
   slowly than in the unmixed one because the temperature near the
   front is higher due to radioactive heating. Note that for the
   unmixed case the outward diffusion 
   of the radioactive energy is clear while for the
   mixed case this effect occurs 
   too early to be noticeable. 
 \label{fig:profilelum}}
\end{center}
\end{figure}

The photosphere, as defined here, is nearly coincident with the outer
edge of the CRW. Therefore it is just the location of the photosphere with
respect to the CRW what sets the value of the photospheric temperature
close to the temperature of hydrogen recombination ($T \sim
5500$~K). However, note that the effective temperature does not necessarily
have a constant value. Since the effective temperature is defined as
$T^4_{\rm eff}= L/\pi \sigma R_{\rm ph}^2$ and the luminosity remains
nearly constant outside the recombination front (see
Figure~\ref{fig:profilelum}), any changes in effective temperature
are related to changes in  photospheric radius.

The CRW divides the object in two distinct regions: (a) an inner zone
which is hot, optically thick and ionized, and (b) an outer 
zone which is relatively cold, optically thin and completely
recombined. Matter in the inner
  zone is opaque: radiative  transfer is 
too inefficient to produce any appreciable flow of energy (this would be
strictly fulfilled if $^{56}$Ni were confined to the innermost layers;
see the discussion below) and the matter cools
down almost adiabatically. On the other hand, the external layers
are transparent and practically do not radiate. Therefore, it is
within the CRW where almost the entire radiant flux is released.  When
matter passes through the CRW,
particles begin to be cooled by radiation, emitting more light than
they absorb, and the radiant flux increases. This way, the radiative
flux emerging from the CRW front carries away internal energy of the
matter that is cooled by the wave. More details on the  properties of
the CRW are given by \citet{1971Ap&SS..10...28G}. 

Note that the bulk of the radiation does not diffuse to the photosphere;
the photosphere instead moves inward, allowing the 
radiation to escape sooner than it would for a photosphere fixed at
the outer boundary of the ejected mass. That is, the recombination
process is responsible for the energy release during this
phase. It should be noted that
most of this energy comes from the energy deposited by the shock wave
and not from the 
recombination itself. In order to test the previous
statement, we ran our code  without including the energy released by 
recombination of ions with electrons. No
appreciable change was found; quantitatively, the differences between
both calculations are less than 0.04 dex during all the evolution. We
have included a figure in the 
electronic edition (Figure~17) showing this comparison. 

Note, however, that in  model \mod\   where we assumed an extended $^{56}$Ni
mixing, the flux of energy inside the CRW is not negligible
(see the left panel of Figure~\ref{fig:profilelum}). 
This is due to the fact that the photosphere meets regions with radioactive
material earlier than in the case where $^{56}$Ni is confined to the innermost
layers. Therefore, the energy deposited by radioactive decay
provides additional power for the LC. A different behavior is seen
when $^{56}$Ni is confined to the innermost layers (inside $2.5$
$M_\odot $; see the right panel of Figure~\ref{fig:profilelum}). In
this case, the statement that the energy flux inside the CRW is very
small is fulfilled at least until day $\sim$110 where
the radiative diffusion of the radioactive decay from the central layers 
begins to dominate. At
earlier times, it is possible to see the outward diffusion of
radioactive energy. Note also that for both models the
luminosity  outside the front is nearly constant.

\begin{figure}
\begin{center}
\includegraphics[angle=-90,scale=.40]{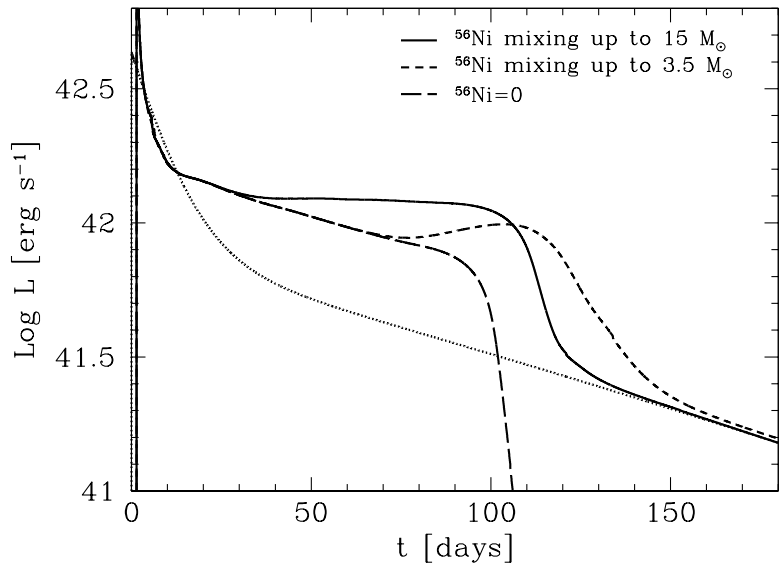}
\caption{Comparison  between bolometric  light curves for extended
   $^{56}$Ni mixing (solid line) and a model with mixing of  $^{56}$Ni  up
  to 3.5 $M_\odot$ (short-dashed line). We also  include the case
  of a model  without  $^{56}$Ni (long-dashed line) in which
  case LC falls abruptly after the plateau phase. The presence of
  $^{56}$Ni extends the plateau and increases the luminosity. This
  is essentially produced when the CRW reaches layers with $^{56}$Ni
  which can thereby power the LC directly. The extended $^{56}$Ni
  mixing reveals this effect earlier on the evolution, which produces
  a flat plateau in concordance with the observations of
  SN~1999em.\label{fig:LCNi}}  
\end{center}
\end{figure}

The assumption of extended mixing was necessary in order to obtain a
plateau as flat as the one observed for
SN~1999em. Figure~\ref{fig:LCNi} shows a comparison of the bolometric
LC for three cases: 1) with extended $^{56}$Ni mixing, 2) no mixing, and
3) without $^{56}$Ni. For case 1), $^{56}$Ni begins to affect the LC
  by day $\sim$35 while for case 2)   
the effect of nickel heating is delayed until day $\sim$75. On the
other hand, in the former case there is a less direct energy
deposition at the  center of the object, and
the plateau declines earlier and steeper to the tail than in case 2). Our
assumption of mixing of $^{56}$Ni into the hydrogen envelope is not
unreasonable as shown in studies of SN~1987A \citep[][among others]
{1988A&A...196..141S,1988ApJ...324..466W,1988ApJ...331..377A,2000ApJ...532.1132B}. However,
 it is important 
to mention that \citet{2007A&A...461..233U} found an excellent agreement
with observations of SN~1999em by confining $^{56}$Ni to the innermost
layers and  good agreement with  observations of SN~1987A assuming
moderate $^{56}$Ni mixing  \citep{1993A&A...270..249U,2004AstL...30..293U} .

\begin{figure}
\begin{center}
\includegraphics[angle=-90,scale=.40]{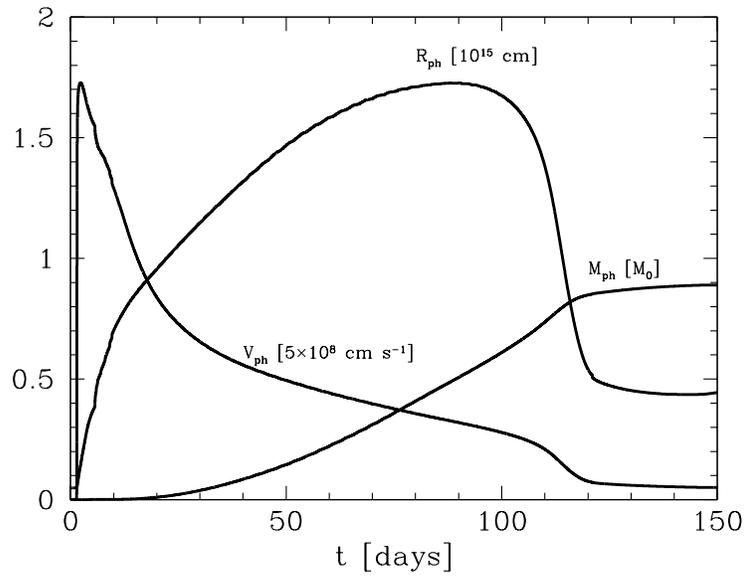}
\caption{Evolution of the photospheric radius ($R_{ph}$) in
    units of $10^{15}$ cm, the photospheric velocity ($V_{ph}$) in
    units of $5 \times 10^{8}$ cm s$^{-1}$, and the mass above
    the photosphere ($M_{ph}$) in units of $M_0$ ($M_0= 19 \,M_\odot$)
    for the  model \mod.} \label{fig:VRM}  
\end{center}
\end{figure}

The evolution of the velocity of matter at the photospheric position
$v_{\rm ph}$, the photospheric radius ($R_{\rm ph}$) and the mass above the
photosphere ($M_{\rm ph}$) are shown in Figure~\ref{fig:VRM}. Initially,
the photospheric velocity evolves rapidly, the photospheric position
follows a linear behavior, and there is very little mass above the
photosphere. Later on, when the recombination sets in, $R_{\rm ph}$
begins to differ noticeably from the linear behavior, the mass above
the photosphere increases, and $v_{\rm ph}$ decreases because it
samples increasingly inner, slower material. Finally, when all the
matter is recombined, $R_{\rm ph}$ and $v_{\rm ph}$ 
sharply turn down and the luminosity undergoes a rapid decrease. Note,
however, that the photospheric radius does not drop immediately.
This is due to the heating caused by radioactive decay which produces
some ionization of the gas. On the other hand, the luminosity undergoes
a rapid decrease to values close to the luminosity of radioactive
decays.  If there was no $^{56}$Ni (case 3) the SN luminosity would
abruptly vanish at
this point, as shown with a long-dashed line in Figure~\ref{fig:LCNi}. 

In conclusion, the CRW has a duration of $\sim$$100$~days. The
luminosity experiences a small change during the CRW propagation.
In order to produce a plateau as flat as that observed for SN~1999em,
we need to invoke Ni mixing in the H-rich envelope. Thus, the plateau
 can be seen as a combination of CRW properties plus some additional
energy provided by radioactivity. Defining, from an empirical point of
view, the plateau phase as the
period of time when the luminosity remains constant within 0.5~mag of
the value at day 50, its duration is $103.5$ days, and the
total energy emitted during this phase is $\sim$$1.12 \times
10^{49}$~erg. Note, that with such definition the plateau
phase includes the final stages of the adiabatic cooling phase as
well.

\subsubsection{Radioactive tail}
\label{sec:tail}
The late behavior of the LC (at $t>130$ days) is dominated by the
energy released from radioactive decay. Without radioactive material, the
luminosity would abruptly vanish when  hydrogen gets completely
recombined as shown in Figure~\ref{fig:LCNi}. Instead, the observed LC
decreases to values close to the instantaneous rate of energy
deposition by the radioactive decay,

\begin{equation}
L\sim 1.43 \times 10^{43} M_{\rm Ni}/M_\odot \,\,\exp(-t/111.3). 
\end{equation}

\noindent The decline of the LC at $t>130$ days is nearly
linear in concordance to the exponential decline of the radioactive
decay law. That is, the diffusion time for optical photons becomes small
enough to radiate away the decay energy instantaneously, while the
gamma-ray optical depth is still sufficiently large in order to
allow a nearly complete local deposition of the decay energy. The
bolometric luminosity in 
this part of the LC is a direct measure of the Ni mass synthesized in the
explosion. 

\subsection{Comparison with other hydrodynamical models of SN 1999em}
\label{sec:compsn99em}

In this Section, we compare our results with those of
previous hydrodynamical studies of SN~1999em. As  
mentioned in Section~\ref{sec:sn99em}, there are two previous
hydrodynamical studies of SN~1999em: one given by
\citet{2005AstL...31..429B} (BBP05 hereafter) and another by
\citet{2007A&A...461..233U} (U07 hereafter). In Table~\ref{tbl-2}, we
summarize the physical parameters obtained for SN~1999em in the three
studies along with the distance, explosion time and
the chemical composition assumed in each model. It is
interesting to note that the three studies employ initial density
distributions and chemical composition distributions of
non-evolutionary models. In our work we assume H and He profiles
very close to those adopted by U07, yet a very different $^{56}$Ni
distribution. While we assume an extended, uniform
mixing of Ni (see Figure~\ref{fig:comp}),  U07 confined 
$^{56}$Ni to the innermost layers (see their Figure~2). In
the case of BBP05, H and He were assumed to be uniformly
mixed throughout the envelope and a radial distribution of $^{56}$Ni up
to $\sim$15 $M_\odot$ was adopted (see their Figure~9). Therefore, our
$^{56}$Ni distribution and the resulting $^{56}$Ni mass are in
better agreement with those of BBP05 than those of U07. We also note
that the mass of the compact remnant --- which is left aside from the
calculations --- is different in all three works. We assume a
compact remnant of 1.4 $M_\odot$ while 
U07 assumed 1.58 $M_\odot$ and BBP05 removed everything within a 
radius of $R_C= 0.1 \, R_\odot$. 

The approaches used in each work are very different. U07 used a 
hydrodynamical code with a one-group approximation for the radiative
transport, including non-LTE treatment of opacities and thermal
emissivity, non thermal ionization, and expansion
opacity. Their calculations, although involving a more sophisticated
 radiative transfer treatment, yielded bolometric light curves as in 
 our code. However, the procedure to produce the observed bolometric
 light curve for SN~1999em was quite 
different in both cases. While U07 based their calculations on the
integration of $UBVRI$ photometry with
a constant value of $0.19$ dex added to take into account the infrared
luminosity, we derived $UBVRIJKLM$ bolometric
luminosities from color-dependent bolometric
corrections. As noted by \citet{2009ApJ...701..200B}, the
  inclusion of the $L$ and $M$ bands in the   
 calculation of the bolometric correction during the tail phase is
 very important and it most likely leads to the 
difference in the nickel mass estimated in this work and in U07.  
BBP05, in turn, used a multi-group hydrodynamical code which allows to
calculate LC in different photometric 
bands. This code also included the expansion opacity
effect. Therefore, BBP05 were able to compare their model with $UBVRI$
LC separately without being affected by the uncertainties in
 the calculation of bolometric luminosities. 
  
As shown in Table~\ref{tbl-2}, the parameters yielded by our
calculations are intermediate between those estimated by BBP05 and
U07. The largest differences are 2.6 $M_\odot$ in mass, 500 $R_\odot$
in radius, 0.3 foes in energy, and $0.045 M_\odot$ in $^{56}$Ni
mass. It is quite satisfactory that our simple 
prescription for the radiative transfer yields physical parameters similar
to those obtained from more sophisticated codes. However, it should be
taken into account that (a) the models and  
methods used in each study are quite different, (b) there are differences
in assumed quantities, such as distance, explosion time, mass cut,
photometry, and (c) there may be a degree of degeneracy among explosion
energy, initial radius and initial mass within each model. Therefore
the present comparison does not help to provide a clear assessment of
the validity of our physical assumptions and of the predictive power
of our model. 

\section{Low Mass SN~1999em}
\label{sec:lowmass}

The mass of the progenitors of SNe~II-P can be derived from hydrodynamical
modeling of light curves and expansion velocities, or from the detection of
the pre-supernova object in archival images of the host galaxy in connection
with stellar evolution models. At the moment three progenitor stars
of SNe~II-P have been firmly detected and there have been negative
detections for 17 other  objects which have led to upper limits of the
progenitor star masses \citep{2009MNRAS.395.1409S}. 
Among these, only three SNe~II-P have masses derived using
hydrodynamical models: SN~1999em, SN~2004et, and SN~2005cs. 
In all of these cases the masses estimated from the hydrodynamical
modeling are   
systematically higher than the values  derived from
the other method. These
discrepancies have been noted previously in the literature
\citep{2008A&A...491..507U,2009arXiv0908.0700S}. For the  particular 
case of SN~1999em, the pre-supernova images give an upper limit of 15
$M_\odot$ for the progenitor star in the ZAMS. It is expected
that the mass of the pre-supernova object should be lower than this due to 
possible mass loss episodes during the evolution of the star. In this Section we
use our hydrodynamical model to explore a low mass range, consistent
with pre-supernova imaging of SN~1999em, in order to test how well we can  
reproduce the observed properties of this object.  Although the
  adoption of non-evolutionary initial models  may introduce an
  uncertainty in the actual progenitor mass (see
  \S~\ref{sec:initialmodel}), it is interesting to study low- and
  high-mass models for this SN in a comparative way.

\begin{figure}
\begin{center}
\includegraphics[angle=-90,scale=.60]{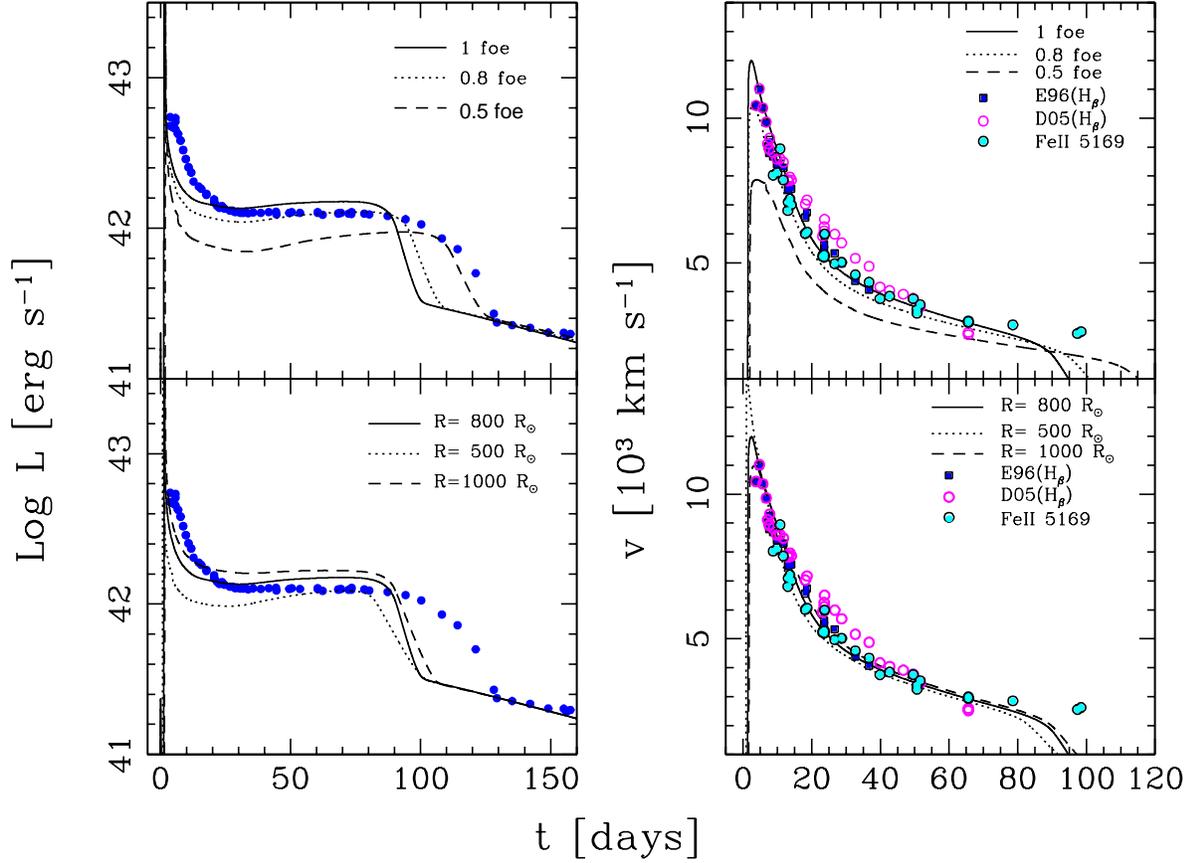}
\caption{Comparison between models and observations of SN~1999em for a
  low value of the pre-supernovae mass of $M = 12 \, M_\odot$ and the
  same value of $^{56}$Ni mass and mixing than for the reference model \mod. 
  {\bf (Left panels):} bolometric LCs. {\bf (Right
    panels):} photospheric velocity evolution. {\bf (Upper
    panels):} models with different explosion energies as
  indicated in the labels, and a fixed value of 
  $R=800 \,R_\odot$. {\bf (Lower panels):} models with different
  initial radii as indicated in the labels, and a fixed value  of $E =
  1$ foe. \label{fig:M12VER}}   
\end{center}
\end{figure}

\begin{figure}
\begin{center}
\includegraphics[angle=-90,scale=.60]{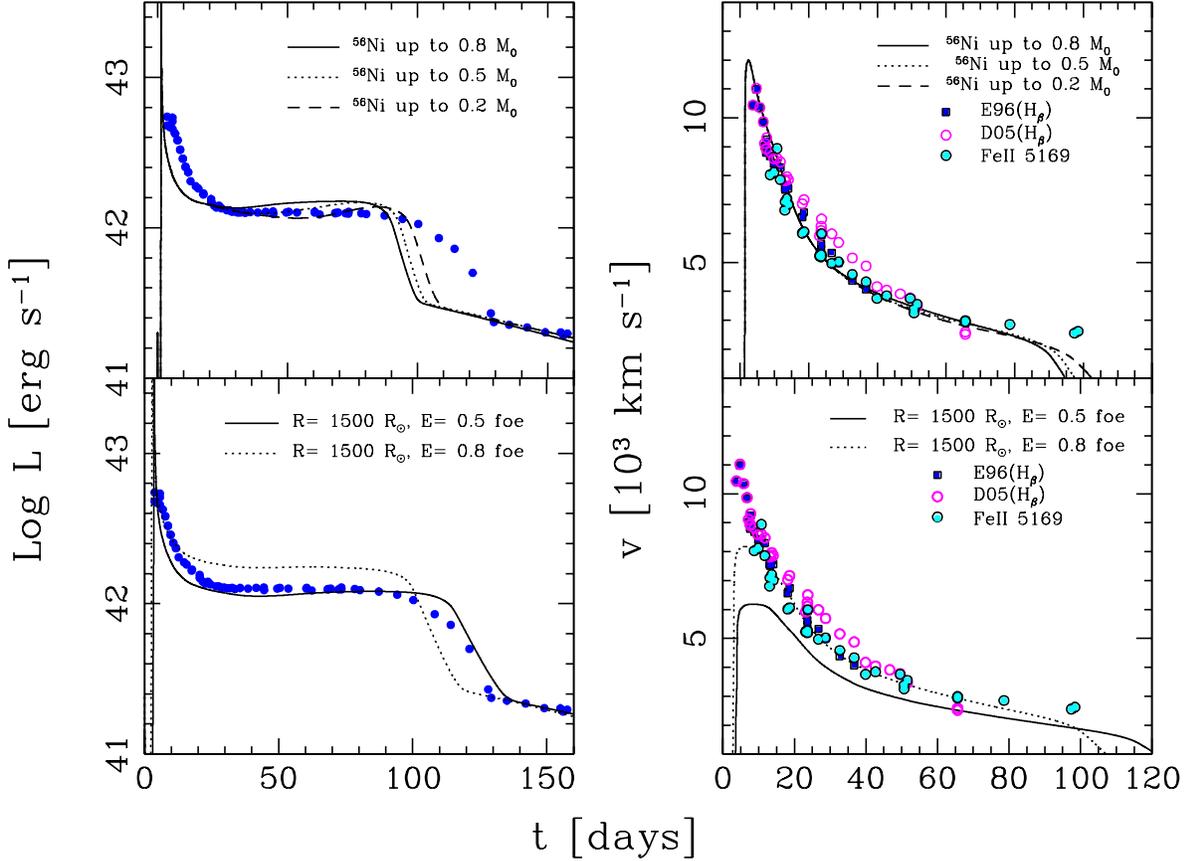}
\caption{Comparison between models and observations of SN~1999em for a
  low value of the pre-supernovae mass of $M = 12 \,M_\odot$ and the
  same value of $^{56}$Ni mass than for the reference model \mod. 
  {\bf (Left panels):} bolometric LCs. {\bf (Right
    panels):} photospheric velocity evolution. {\bf (Upper
    panels):} models with different mixing of
    $^{56}$Ni, $E=1$ foe and $R=800 \,R_\odot$. Note that the degree
  of $^{56}$Ni  is indicated in each figure as a fraction of the
  initial mass of the 
  model. {\bf (Lower panels):} models with an initial
  radius of $R= 1500 \, R_\odot$ and two different values for  the
  explosion energy as indicated (corresponding to models m12r15e08ni56
  and m12r15e05ni56; see Table~\ref{tbl-3}).  Model m12r15e05ni56
reproduces very well the bolometric light curve of SN~1999em, it while fails
to reproduce the photospheric velocity evolution.\label{fig:M12VNi}} 
\end{center}
\end{figure}

\begin{figure}
\begin{center}
\includegraphics[angle=-90,scale=.60]{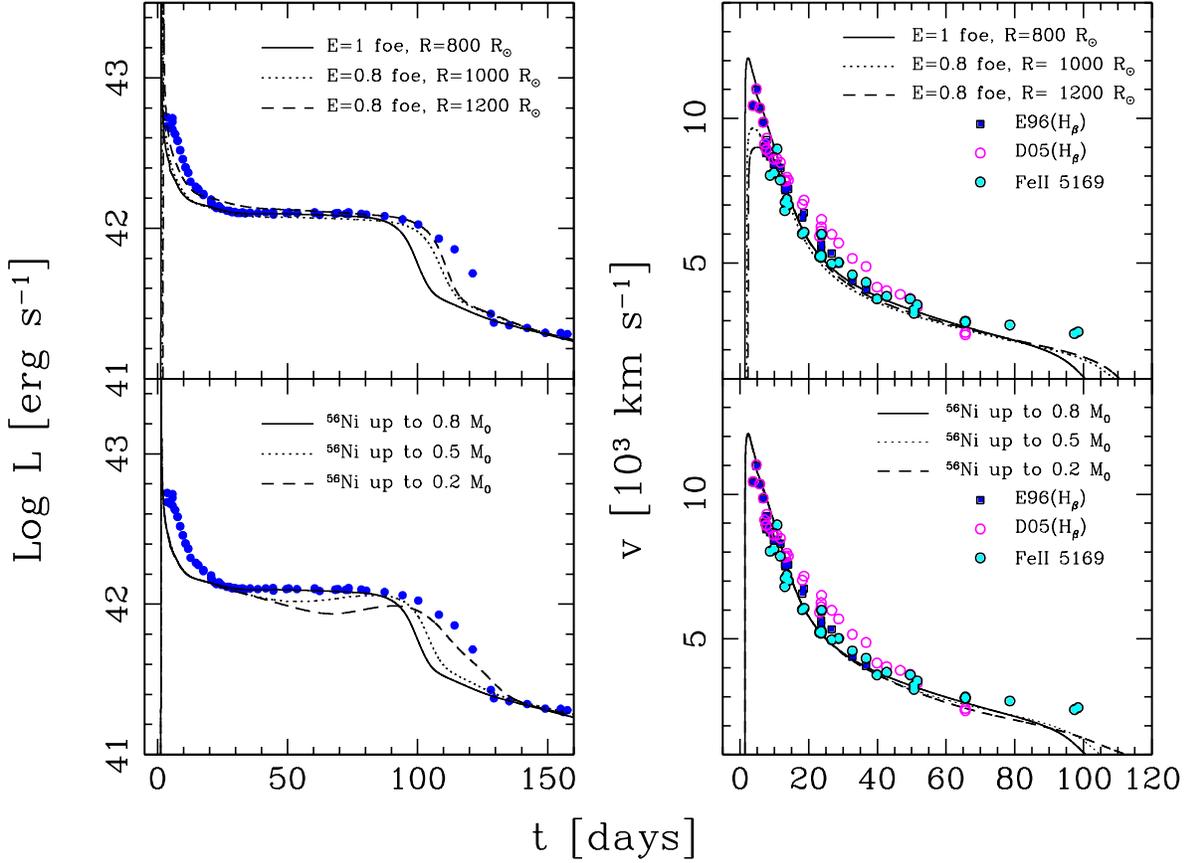}
\caption{Comparison between models (lines) and observations
    (points) of SN~1999em for a
  low value of the pre-supernova mass of $M = 14 \,M_\odot$ and the
  same mass of $^{56}$Ni as for the reference model \mod. 
  {\bf (Left panels):} bolometric LCs. {\bf (Right
    panels):} photospheric velocity evolution. {\bf (Upper
    panels):} models {\sc m14r8e1ni56}, {\sc m14r10e08ni56} and {\sc
    m14r12e08ni56} (see Table~\ref{tbl-3}). {\bf (Lower 
  panels):} models with different mixing of 
    $^{56}$Ni, $E=1$ foe and $R=800 \,R_\odot$. Note that the degree
  of $^{56}$Ni  is indicated in each panel as a fraction of the
  initial mass of the model. \label{fig:M14VP1}} 
\end{center}
\end{figure}

\begin{figure}
\begin{center}
\includegraphics[angle=-90,scale=.60]{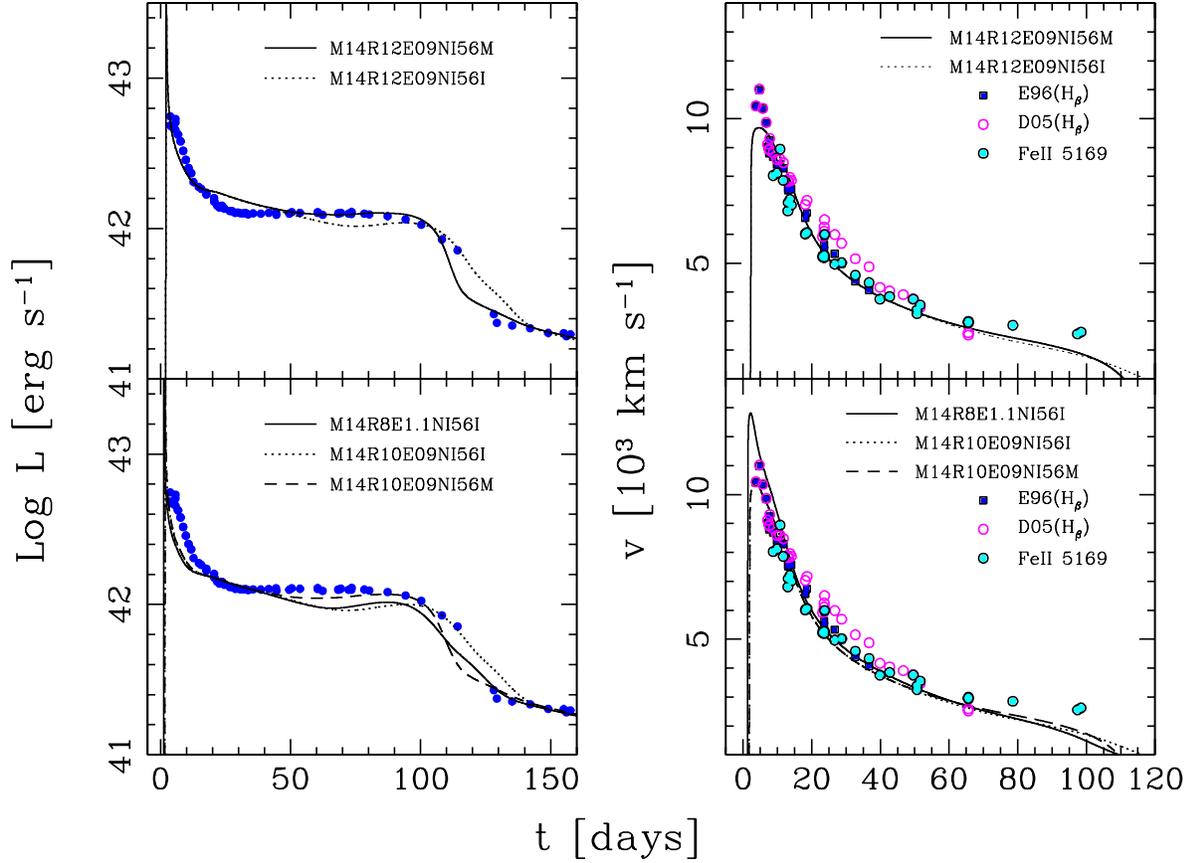}
\caption{Comparison between models (lines) and observations
    (points) of SN~1999em for a
  low value of the pre-supernova mass of $M = 14 \,M_\odot$ and the
  same mass of $^{56}$Ni as for the reference model \mod. 
  {\bf (Left panels):} bolometric LCs. {\bf (Right
    panels):} photospheric velocity evolution. {\bf (Upper
    panels):} models {\sc m14r8e1.1ni56I}, {\sc m14r10e09ni56I} and {\sc
    m14r10e09ni56M}. {\bf (Lower 
  panels):} models {\sc m14r12e09ni56M} and {\sc m14r12e09ni56I} (see
  Table~\ref{tbl-3}). \label{fig:M14VP2}}  
\end{center}
\end{figure}

We calculate several models using two different values of the
pre-supernova mass: $M= 12 \, M_\odot$ (see Figures~\ref{fig:M12VER} and
\ref{fig:M12VNi}), and $M= 14 \, M_\odot$ (see
Figures~\ref{fig:M14VP1} and \ref{fig:M14VP2}). In all 
Figures, we show the bolometric light curves
and  photospheric velocities ($v_{ph}$)  for
different values of the injected energy ($E$), initial radius ($R$),
and degree of $^{56}$Ni mixing. Table~\ref{tbl-3} gives a summary of
the parameters used. We remark that in all of these models, we
have assumed the same value for the $^{56}$Ni mass as that of model
\mod, i.e. $0.056 M_\odot$,  because this value is required to
  fit  the tail of the LC. Also note that the degree of $^{56}$Ni
  mixing is taken 
as a fraction of the initial mass of the object in order to compare
the different degrees of mixing in a consistent way when models 
with different initial mass are used. For example, a model with a mixing of
$^{56}$Ni up to $0.8 M_0$ has an equivalent degree of $^{56}$Ni mixing
as that of model \mod.

For the case of  12 $M_\odot$, we have calculated nine
models. We first analyze the effect on the results of the variation of
one parameter while keeping the other parameters fixed. In the upper panel
of Figure~\ref{fig:M12VER} we show the 
effect on the LC and $v_{ph}$ of the variation of $E$ for a
model with $R = 800 \, R_\odot$ and $^{56}$Ni mixing up to $0.8 \, M_0$. In
the bottom panels of Figure~\ref{fig:M12VER}, we show the effect of varing
$R$, for a model with $E=1$ foe and for the same
$^{56}$Ni mixing. The sensitivity of the LC and $v_{ph}$  on
the extent of  $^{56}$Ni mixing is shown in the upper panel of
Figure~\ref{fig:M12VNi} for a model with $E=1$ foe and $R = 800 \, R_\odot$.

An examination of these Figures yields the following conclusions: 1)
higher injected energy produces higher luminosity and a shorter
plateau, 2) larger initial radius produces higher luminosity and a longer
plateau, 3) more extended $^{56}$Ni mixing produces higher
luminosity and a shorter plateau, 4) $^{56}$Ni mixing and initial
radius have a small 
effect on the $v_{ph}$ evolution as compared with that of the
explosion energy, and 5) as expected, there is no effect of any of
these parameters on the tail luminosity.
Note that the sensitivity of the LC on $E$, $R$, and
$^{56}$Ni mixing is in qualitative  concordance with analytic studies
by \citet{1980ApJ...237..541A}, \citet{1993ApJ...414..712P}, and previous
numerical studies by \citet{1983Ap&SS..89...89L,1985SvAL...11..145L}
and more recently by 
\citet{2004ApJ...617.1233Y} and \citet{2007A&A...461..233U}.

Although we have considered several values of $E$, $R$, and $^{56}$Ni mixing,
neither of the models with  $M= 12 \,  M_\odot$  gives a good
representation of the 
observations of SN~1999em. Note that we have  calculated also  two other
models, {\sc m12r15e05ni56} and {\sc m12r15e08ni56}, adopting even
higher values of the 
initial radius. These models are shown in the bottom panel of
Figure~\ref{fig:M12VNi}. Despite the good representation of the LC
provided by model {\sc m12r15e05ni56}, it 
fails to reproduce the observed  photospheric velocities. We conclude
that it is not possible 
to reproduce the observations of SN~1999em using models with a
  pre-supernova mass of $M = 12 \,  M_\odot$. Clearly, masses lower
  than this value would also fail to match the observations.  

For the case of 14 $M_\odot$, we have calculated ten models for
different combinations of $R$, $E$ and $^{56}$Ni
mixing. Specifically, we used three different values of the initial
radius: $R = 800$, $1000$, and $1200 \,R_\odot$, four values of
explosion energy: $E = 1.1$, $1.0$, $0.9$, and $0.8$ foe, and three
different degrees of $^{56}$Ni mixing: $0.2$, $0.5$, and $0.8 \, M_0$ (see
Table~\ref{tbl-3}). Figures~\ref{fig:M14VP1} 
and \ref{fig:M14VP2} show the results of these models. At first 
sight we find that these models agree better with the observations
than the 12 $M_\odot$ ones. Therefore the case of 14 $M_\odot$
deserves a more detailed analysis. As described below, we
chose the grid of parameters based on the  
known effects of each physical parameter on the light curves and
velocities, and trying to reach the best possible agreement with the
observations. As noted in Section~\ref{sec:hydro}, in order to assess
the validity  
of models we will focus on how well they reproduce the observed
plateau luminosity and length, and the early photospheric velocities, but we
will disregard differences in the early LC before the plateau phase
where the uncertainties in both models and data are the largest.

Our reference model is shown with a solid line in the upper
  panels of Figure~\ref{fig:M14VP1} and has the following parameters:
  $E = 1$ foe, $R=800\,R_\odot$, and $0.8 \, M_0$ of $^{56}$Ni
  mixing. We note that this reference model has the same initial radius and
  $^{56}$Ni mixing as model \mod\ for 19 $M_\odot$. The energy has
  been reduced in order to compensate the effect on the luminosity of
  using a lower mass. As shown in Figure~\ref{fig:M14VP1} the reference model
  produces the correct plateau luminosity and photospheric velocities
  although the plateau duration is too short as compared with the
  data. With the aim of remedying this situation while keeping 
  the mass fixed we invoked lower energies and larger radii (dashed
  and dotted lines). While this served to improve the issue of the
  plateau length, the 
  comparison with photospheric velocities at early times became
  poorer as  expected due to the lower energies. We therefore
   consider these models to be unlikely. In a
  further attempt to find a good 14 $M_\odot$ model, we decided to
  vary the mixing of $^{56}$Ni while keeping the other parameters as in
  the reference model. The results are shown in the lower panels of 
  Figure~\ref{fig:M14VP1}. We now note that while a reduction of the
  $^{56}$Ni mixing serves to increase the length of the plateau, the
  shape of the LC and the plateau luminosity drift away from the
  observations. Thus, we can also discard these models.

Based on the above observations we tested other parameter
combinations, as shown in Figure~\ref{fig:M14VP2}. In the upper panels
we show the tests of models with slightly smaller energies and larger
radii as compared with the reference model, and two different
degrees of $^{56}$Ni mixing. While the match of the LC for these
models is satisfactory, the problem with the early-time velocities
reappears, so we discard these models. 
In the lower panels of Figure~\ref{fig:M14VP2} we show further combinations of
parameters. Here we obtain an improvement in the LC while not
compromising the agreement in the velocities. Among these models the
one called {\sc m14r10e09ni56m} provides the best match  to the LC and
velocity data. We therefore conclude that the 14 $M_\odot$ scenario
for SN~1999em cannot be ruled out, although the comparison with the
data is not as good as that of the 19 $M_\odot$ model (see
Figure~\ref{fig:LC} and \ref{fig:velo}).

\section{CONCLUSIONS}
\label{sec:conclusion}

We have developed a one-dimensional, flux-limited Lagrangian
hydrodynamical code useful for modeling bolometric light curves and
the photospheric velocity evolution of SNe~II-P. The performance of the
code was examined by its application to one of the best observed SNe~II-P,
SN~1999em, obtaining a very good agreement with the observations with the
following physical parameters $E=1.25$ foe, $M= 19 \,M_\odot$, $R= 800
\,R_\odot$, and $M_{\mathrm {Ni}}=0.056 \,M_\odot$. In our analysis, we
found that an extended mixing of $^{56}$Ni is needed in order to
reproduce a plateau as flat as that shown by the observations of
SN~1999em. We note that the plateau phase, at least in the case of
this SN, is powered by the energy deposited by the shock wave and
released by the recombination process, plus some extra energy
deposited by the radioactive material.

Our model for SN~1999em required extensive mixing of $^{56}$Ni into
the envelope. This has also been strongly suggested by the modeling of
the observations of SN~1987A \citep[see][among
  others]{1988A&A...196..141S,1988ApJ...324..466W,1988ApJ...331..377A,2000ApJ...532.1132B}.  
In addition, \citet{2005AstL...31..429B} have 
also invoked  extensive $^{56}$Ni mixing to model SN~1999em.
\citet{2004ApJ...617.1233Y} explored the effect on the LCs of SNe~II of
the variation of several parameters and found, in concordance with our
study, that between 50 and 120 days, the LC is powered by the energy
deposited by the shock wave in combination with energy deposited by gamma rays
when an extended mixing of $^{56}$Ni is used.

We compare our results for SN~1999em with two 
hydrodinamical studies of this object previously given by BBP05 and
U07. Although there are differences in the physical parameters used in
each work we believe that the results are remarkably
consistent considering the differences in distance, explosion time,
mass cut, the observational data employed, and the very different
prescriptions for the radiative transfer. Note that the
physical parameters used in our calculations are intermediate between
those estimated by BBP05 and U07.  Although it is satisfactory that our
code provides good fits to the observations with physical
parameters similar to those obtained from more sophisticated
calculations, we note that given the differences among the three
approaches, the comparison does not help to assess the validity of our
assumptions nor the predictive power of the methods.

We have also explored the feasibility of fitting the
observations of SN~1999em with low mass models, consistent with  
the upper limits obtained through pre-supernova imaging in connection
with stellar evolution models. We were not able to find a set of physical
parameters that reproduce well the observations when assuming  
pre-supernova masses of $\leq$ 12  $M_\odot$. On the other hand, models with 14
$M_\odot$ cannot be fully discarded. Specifically, we found one model 
with the following parameters: $E = 0.9$ foe, $R=1000\,R_\odot$, and
$0.8 \, M_0$ of $^{56}$Ni mixing which provides a reasonable fit to
the observations, although not as good as our favorite model of 19
$M_\odot$.  We remark that even if the exact values of the
  progenitor mass  may be affected by the choice of parametric
  initial models, this comparative analysis favors models with 
  total  masses larger than 14 $M_{\odot}$.

The fact that models with different masses  agree
reasonably well with the observations tells us that there
is a degree of degeneracy among the physical parameters which must be
quantified if one wants to assess the precision to which such
parameters can be determined using hydrodynamical
calculations. We plan to continue investigating this issue with our code.

Additional information such as modeling of spectra can help
discriminating among possible scenarios. On the other hand, one way to
  reduce the number of free 
parameters would be to use initial models from stellar evolution
calculations. However, 
 the use of evolutionary initial models
introduces unobserved features in the resulting light curves
\citep[see Section~\ref{sec:initialmodel} and
  also][]{2008A&A...491..507U} probably caused by sharp boundaries
between layers with different chemical  
compositions and/or the steep density jump between the helium core and the 
hydrogen envelope. With our parametric density and chemical profiles
we were able to smooth the transitions and thereby obtain smooth
LC.  The disadvantage of this approach is that parametric initial
  models are less directly connected with actual properties of the
  progenitor star.

As  has been previously proposed, we suggest that a 
possible reason for the need of invoking ad-hoc initial models is the
fact that the calculations of the shock propagation are done in one
dimension. Therefore, known effects that cause mixing, such as
Rayleigh-Taylor (RT) instabilities, cannot be taken into account. Note
that even if RT instabilities produce mixing behind the shock, the
effect that is 
observed in our modeled light curves is related to the structure that
is {\em left behind}  the passage of the shock once it has reached the
surface. Such structure is revealed as the recombination front
recedes into the ejecta. Therefore, we think that it is important to
attempt to quantify the effect of the RT instability on the mixing and
density structure in one dimension in order to check if this effect
can explain the type of initial distributions and mixing assumed in
our calculations.    

The remarkably good agreement with the observation of SN~1999em and the
consistency found with previous hydrodynamical studies of this SN
using more sophisticated codes gives us confidence in our attempt
to model SNe~II-P, at least during the plateau phase and early
evolution of the radioactive tail phase, despite the
simplifications assumed in our calculation. This work is
   the starting point for the analysis of the physical
  parameters of our sample of 33 SNe~II-P
 that will be done in a forthcoming paper with the aim of improving our
 understanding of these objects.

\acknowledgments
The authors gratefully acknowledges the helpful conversations
  with Sergei Blinnikov, Ken'ichi Nomoto, Nozomu Tominaga and 	
Alejandro Clocchiatti. A special gratitude to Gast\'on Folatelli for
his support and generous revision of the text.
MB acknowledges support from MECESUP UCH0118 program  and the
  World Premier International Research Center Initiative, MEXT,
Japan. OGB is member of the Carrera del Investigador 
Cient\'{\i}fico, Comisi\'on de Investigaciones Cient\'{\i}ficas de la
Provincia de Buenos Aires, Argentina. MH obtained 
support from proyecto FONDECYT (grant 1060808), Millennium Center for
Supernova Science through grant P06-045-F, Centro de Astrof\'\i sica
FONDAP 15010003, and Center of Excellence in Astrophysics and
Associated Technologies (PFB 06). 

\clearpage

\clearpage

\clearpage
\begin{deluxetable}{lcccccccccc}
\rotate
\tablecaption{Properties of model \mod\, at selected time of their evolution.
  \label{tbl-1}}
\tablewidth{0pt}
\tablehead{
\colhead{}  &  \colhead{$t$}  &\colhead{$\log(L_{\mathrm{bol}})$}& \colhead{$\log(T_{\mathrm{eff}})$}&
\colhead{$\log(T_{\mathrm{ph}})$}& \colhead{$R_{\mathrm{ph}}$}& \colhead{$v_{\mathrm{ph}}$} &
\colhead{$E_{\mathrm{rad}}/E_0$\tablenotemark{a}}& 
\colhead{$E_{\mathrm{K}}/E_0$\tablenotemark{a}}\\
\colhead{Phase}  &\colhead{[days]} & \colhead{[erg~s$^{-1}$]} & \colhead{[K]} & \colhead{[K]}&
\colhead{[10$^{14}$cm]} & \colhead{[10$^{8}$cm~s$^{-1}$ ]} & \colhead{(\%)} &
\colhead{(\%)}  }  
\startdata 
Peak       &    1.5     &  44.5   &  5.05 & 5.03  &  0.54  & 3.36 & 0.04 & 72  \\
Adiabatic Cooling  &    5  &  42.4   &  4.10 & 4.05  &  3.8   & 11.5 & 0.2  & 93  \\
Plateau    &    50      &  42.1   &  3.73 & 3.77  &  14.7  & 3.47 & 0.65 & 99  \\
Transition &    114     &  41.7   &  3.74 & 3.74  &  10    & 1   & 1.2  & 99.7 \\
\enddata
\tablenotetext{a}{$E_{rad}$ is defined by $\int_0^t L_{bol} \, dt$ and
  $E_K$ is the total kinetic energy in the mass motions. In the table
  we show these quantities normalized to the initial injected  energy
  of  model \mod, $E_0=1.25$ foe.}
\end{deluxetable}

\begin{deluxetable}{lccccccccc}
\rotate
\tablecaption{Comparison between physical parameters for SN 1999em
  from three different hydrodynamical codes
  \label{tbl-2}}
\tablewidth{0pt}
\tablehead{
\colhead{}  &\colhead{$D$}& \colhead{$t_{0}$}&
\colhead{}& \colhead{}& \colhead{$E$} & \colhead{$M$}&
\colhead{$R$}& \colhead{$M_{\mathrm {Ni}}$} & \colhead{Ni mixing}\\
\colhead{Code} & \colhead{[Mpc] } & \colhead{[JD-2451000]} & \colhead{$X_{\mathrm{sup}}$}&
\colhead{$Z$} & \colhead{[foe]} & \colhead{[$M_\odot$]} &
\colhead{[$R_\odot$]}  & \colhead{[$M_\odot$]} & \colhead{[$M_\odot$]} }  
\startdata 
This work           &  11.7   &  477.90  & 0.735 &  0.02  & 1.25 & 19    & 800  &
0.056 &  $\sim$15\\
BBP05\tablenotemark{a}&  12  & 468.90  &  0.7   &  0.004  & 1   & 18    & 1000 &
0.06 & $\sim$15 \\
U07\tablenotemark{b} &  11.7  &  476.90 & 0.735 &  0.017 & 1.3$\pm$ 0.1
& 20.58$\pm$1.2 & 500$\pm$ 200  &
0.036$\pm$0.009 &  $\sim$2.5 \\
\enddata

\tablenotetext{a}{\citet{2005AstL...31..429B}} 
\tablenotetext{b}{\citet{2007A&A...461..233U}}
\end{deluxetable}

\begin{deluxetable}{lcccc}
\tablecaption{ Model Parameters\label{tbl-3}}
\tablewidth{0pt}

\tablehead{
\colhead{}  &\colhead{Mass}& \colhead{Radius} & \colhead{Energy}& \colhead{Ni} \\
\colhead{Model} & \colhead{[$M_\odot$]} & \colhead{[$R_\odot$]} &\colhead{[foe]} &\colhead{mixing\tablenotemark{a}} }  
\startdata
{\sc m12r8e1ni56}      &  12    &  800   &   1      &   0.8    \\
{\sc m12r8e08ni56}     &  12    &  800   &   0.8    &   0.8    \\
{\sc m12r8e05ni56}     &  12    &  800   &   0.5    &   0.8    \\
{\sc m12r5e1ni56}      &  12    &  500   &   1      &   0.8    \\
{\sc m12r10e1ni56}     &  12    &  1000  &   1      &   0.8    \\
{\sc m12r8e1ni56m}     &  12    &  800   &   1      &   0.5    \\
{\sc m12r8e1ni56in}    &  12    &  800   &   1      &   0.2    \\
{\sc m12r15e05ni56}    &  12    &  1500  &   0.5    &   0.8    \\
{\sc m12r15e08ni56}    &  12    &  1500  &   0.8    &   0.8    \\
{\sc m14r8e1ni56}      &  14    &  800   &   1      &   0.8    \\
{\sc m14r10e08ni56}    &  14    &  1000  &   0.8    &   0.8    \\
{\sc m14r12e08ni56}    &  14    &  1200  &   0.8    &   0.8    \\
{\sc m14r8e1ni56m}     &  14    &  800   &   1      &   0.5    \\
{\sc m14r8e1ni56i}     &  14    &  800   &   1      &   0.2    \\
{\sc m14r8e1.1ni56i}   &  14    &  800   &   1.1    &   0.2    \\
{\sc m14r10e09ni56m}   &  14    &  1000  &   0.9    &   0.5    \\
{\sc m14r10e09ni56i}   &  14    &  1000  &   0.9    &   0.2    \\
{\sc m14r12e09ni56m}   &  14    &  1200  &   0.9    &   0.5    \\
{\sc m14r12e09ni56i}   &  14    &  1200  &   0.9    &   0.2    \\
\enddata
\tablenotetext{a}{The degree of $^{56}$Ni mixing is given as a
    fraction of the initial mass of the model ($M_0$)}
\end{deluxetable}

\end{document}